\documentclass[aps,11pt,preprintnumbers,nofootinbib,floatfix]{revtex4}

\usepackage{graphicx}% Include figure files
\usepackage{epsfig}
\usepackage{dcolumn}% Align table columns on decimal point
\usepackage{subfigure}
\usepackage{multirow}
\usepackage{amsmath}
\usepackage{amssymb}

\def\be{\begin{equation}}
\def\ee{\end{equation}}
\def\bea{\begin{eqnarray}}
\def\eea{\end{eqnarray}}

\begin{document}

\title{17 MeV Atomki anomaly from short-distance structure of spacetime}

%%%% To generate auto affiliation numbers please use \author{}\affil{} command

\author{Cao H. Nam}
\email{nam.caohoang@phenikaa-uni.edu.vn}  
\affiliation{Phenikaa Institute for Advanced Study and Faculty of Basic Science, Phenikaa University, To Huu, Yen Nghia, Ha Dong, Hanoi 100000, Vietnam}
\date{\today}

\begin{abstract}%
An anomaly observed recently in the $^8$Be nuclear transition by the Atomki collaboration hints at a weakly-coupled, light new gauge boson with a mass of about $17$ MeV. In this paper, we propose that this new gauge boson comes from a short-distance structure of the spacetime, rather than from an extension of the Standard Model through adding an extra $\mathrm{U}(1)$ gauge symmetry. The dominant contribution to the relevant matrix element of the $^8$Be nuclear transition is given by the axial couplings. In order to account for the $^8$Be anomaly and satisfy the current experimental constraints, the coupling constant of the new gauge boson should be about $\mathcal{O}(10^{-4}-10^{-5})$. Our theoretical model allows understanding the origin of the smallness of the coupling constant, which is still missing or being incompletely understood in the models at which the new gauge boson has the axial couplings.
\end{abstract}

\maketitle

\section{Introduction}

There have been the experimental evidences which indicate new physics beyond the Standard Model (SM), such as the tiny masses of the neutrinos and their mixing, dark matter, or matter-antimatter asymmetry. The simplest extension of the SM is to add new Abelian gauge forces corresponding to the $\mathrm{U}(1)$ symmetry groups. No signal for new Abelian gauge forces has so far been detected at the high energy colliders such as the LHC as well as they have been tested indirectly through the high accurate measurements. Since these Abelian gauge bosons should be heavy with the masses (much) larger than the electroweak scale. On the contrary, they may be light but are coupled very weakly to the SM particles, which allows them to be hidden under the searches at the low energy collider experiments such as BABAR or BELLE.

The popular approach to add an extra $\mathrm{U}(1)$ symmetry is through the extension of the SM gauge symmetry group. There, the additional $\mathrm{U}(1)$ symmetry could arise from grand unified theories (GUTs) \cite{Langacker1981,Rosner1982a,Rosner1982b,Rosner1984,Rosner1986,Rizzo1989}, from the left-right symmetric models \cite{Mohapatra1975,Mohapatra1980,Mohapatra1981}, from various simple
extra $\mathrm{U}(1)$ gauge symmetries \cite{Tait2004,Nath2004a,Nath2004b,Nath2005,
Pokorski2006,Nath2007,Khalil2008,Gopalakrishna2008,Masiero2008,Khalil2010,Orikasa2009,Burell2012,HSLee2012,
HSLee2013,Sugiyama2014,TCYuan2014,Takahashi2015,Das2016,Okada2017,Das2017,Zakeri2017,Chao2018,Raut2018,Das2018,SOkada2018,
Okada2018a,Okada2018b,Dhargyal2018,Das2019}. However, there is an alternative possibility that the additional $\mathrm{U}(1)$ symmetry arises from a more fundamental structure of the spacetime. In this sense, this additional $\mathrm{U}(1)$ symmetry is actually emergent, rather than being considered as a fundamental gauge symmetry. In this direction, starting fundamentally from a generally covariant theory which consists the fields propagating dynamically in a five-dimensional fiber bundle spacetime $M_5$ and respecting for the SM gauge symmetry group $\mathrm{SU}(3)_C\otimes\mathrm{SU}(2)_L\otimes\mathrm{U}(1)_Y$, Ref. \cite{Nam2019a} obtained a $\mathrm{U}(1)$ extension of the SM in the four-dimensional effective spacetime. 

The Atomki Collaboration has recently reported an excess of the electron-positron pairs, with a high statistical significance of $6.8\sigma$, produced in the $^8\text{Be}^*(1^+)\rightarrow{^8\textrm{Be}}(0^+)+e^+e^-$ transitions \cite{Krasznahorkay2016} (see also Refs. \cite{Ketel2016,Krasznahorkay2017a,Krasznahorkay2017b,Krasznahorkay2017c,Krasznahorkay2018}
). The $^8$Be anomaly can be interpreted by a weakly-coupled, light new gauge boson produced on-shell in the decay of the excited state $^8\text{Be}^*$ and subsequently decaying into the electron-positron pairs. 
The best fit to the mass of the new gauge boson is $16.70\pm0.35(\text{sta})\pm0.5(\text{sys})$ MeV. Various models have been proposed to explain the $^8$Be anomaly in different $\mathrm{U}(1)$ gauge symmetries \cite{Feng2016,Jia2016,Ellwanger2016,Kitahara2017,Feng2017,Seto2017,Chen-Liu2017,Gu-He2017,Liang2017,Kozaczuk2017,Rose2017,Chiang2017,Jiang2018,Rose2019}. In these studies, the $\mathrm{U}(1)$ gauge symmetry corresponding to the new gauge boson is introduced through the extension of the SM gauge symmetry group.

The goal of the present paper is to investigate whether the weakly-coupled, light new gauge boson accounting for the $^8$Be anomaly can come from the more fundamental structure of the spacetime. The new gauge boson in this scenario has both axial and vector couplings to the quarks. (Note that, it is not easy to construct a model with only vector couplings to the quarks, because the coupling constant to the quarks is relatively large to explain the $^8$Be anomaly, which is thus difficult to evade various experimental bounds.) As indicated in Refs. \cite{Kozaczuk2017,Rose2017,Rose2019}, in such models, the coupling constant of the new gauge boson is about $\mathcal{O}(10^{-4}-10^{-5})$ which the origin of this smallness is still missing or being incompletely understood. Our scenario is based on Ref. \cite{Nam2019a} with the suitable modifications, where the coupling constant of the new gauge boson is proportional to the ratio between the inverse radius $1/R$ of the extra space and the reduced four-dimensional Planck scale. In Ref. \cite{Nam2019a}, it is indicated that for producing the active neutrinos of the sub-eV mass scale, the inverse radius $1/R$ is about $10^{14}$ GeV with the Yukawa coupling $h^{\nu}$ to be the order of the Yukawa coupling of the top quark. This value of the inverse radius $1/R$ thus predicts the gauge coupling $g_{_X}\sim\mathcal{O}(10^{-5})$ which is consistent to the $^8$Be anomaly.

This paper is organized as follows. In Sec. \ref{model}, we introduce a model at which a new $\mathrm{U}(1)_X$ gauge symmetry, corresponding to a new neutral gauge boson, is emerged from the more fundamental structure of the spacetime. We determine the charges of the SM particles under this $\mathrm{U}(1)_X$ group and the coupling terms. In Sec. \ref{Beanomaly}, we show that our model can provide an interpretation of the $^8$Be anomaly consistent with the current experimental bounds. The last section is devoted to our conclusions.

\section{\label{model} Model Setup}

In this section, we will review briefly the model proposed in Ref. \cite{Nam2019a}, with including the modifications to have a consistent model to explain the $^8$Be anomaly. For the more details of this model we refer the reader to Ref. \cite{Nam2019a}.

\subsection{Description of spacetime}

It was proposed that the spacetime at more fundamental level is a five-dimensional fiber bundle $M_5$ whose base manifold and fiber are the four-dimensional manifold $M_4$ of the Lorentz signature and the Lie group manifold $\mathrm{U}(1)$, respectively. In general, one local region of the spacetime $M_5$ looks like a product $V_i\times\mathrm{U}(1)$ with $V_i$ to be a local region (an open subset) of $M_4$. As a result, each $V_i\times\mathrm{U}(1)$ enables to assign the local coordinates for a point in the spacetime $M_5$ as, $(x^\mu,e^{i\theta})$, where $\{x^\mu\}\in V_i\subset M_4$ and $e^{i\theta}\in\mathrm{U}(1)$ with $\theta$ to be dimensionless real parameter. Two local coordinate systems $V_i\times\mathrm{U}(1)$ and $V_j\times\mathrm{U}(1)$, which are assigned to a same spacetime point, are related together by the general coordinate transformation as
\begin{eqnarray}
x^\mu&\longrightarrow&x'^\mu=x'^\mu(x),\nonumber\\
e^{i\theta}&\longrightarrow&e^{i\theta'}=h(x)e^{i\theta},\ \
\textrm{or}\ \
\theta\longrightarrow\theta'=\theta+\alpha(x).\label{ba-gct}
\end{eqnarray}
The theory is covariant with respect to this general coordinate transformation. It should be noted here that if the base manifold $M_4$ is Minkowski-flat (meaning that the usual four-dimensional gravitation is not taken into account) the general coordinate transformation (\ref{ba-gct}) becomes
\begin{eqnarray}
x^\mu&\longrightarrow&x'^\mu=x^\mu,\nonumber\\
\theta&\longrightarrow&\theta'=\theta+\alpha(x).\label{gct}
\end{eqnarray}

In preparation for proceeding, we would like to present some important properties of the spacetime $M_5$.
The tangent space $T_pM_5$ at a point $p\in M_5$ is always decomposed into a direct sum of four-dimensional horizontal tangent subspace $H_pM_5$ and one-dimensional vertical tangent subspace $V_pM_5$ as
\begin{equation}
T_pM_5=H_pM_5\oplus V_pM_5,
\end{equation}
without reference to the local coordinate system \cite{Nakahara}. Respectively, the tangent subspaces $H_pM_5$ and $V_pM_5$ are spanned by the following covariant bases
\begin{eqnarray}
\left\{\frac{\partial}{\partial x^\mu}-g_{_X}X_\mu\frac{\partial}{\partial\theta}\equiv\hat{\partial}_\mu\right\},\ \ \ \ \ \left\{\frac{\partial}{\partial\theta}\equiv\partial_\theta\right\},
\end{eqnarray}
where the gauge field $X_\mu$ transforms under the general coordinate transformation (\ref{ba-gct}) as
\begin{eqnarray}
X_\mu\longrightarrow
X'_\mu&=&\frac{\partial x^\nu}{\partial x'^\mu}\left(X_\nu-\frac{1}{g_{_X}}\partial_\nu\alpha(x)\right).\label{B-gautrn}
\end{eqnarray}
and $g_{_X}$ is the gauge coupling. It is easily to check that these bases transform under the general
coordinate transformation (\ref{ba-gct}) as
\begin{eqnarray}
\hat{\partial}_\mu\longrightarrow\hat{\partial}'_\mu&=&\frac{\partial}{\partial x'^\mu}-g_{_X}X'_\mu\frac{\partial}{\partial\theta'},\nonumber\\
&=&\frac{\partial x^\nu}{\partial x'^\mu}\frac{\partial}{\partial x^\nu}+\frac{\partial\theta}{\partial x'^\mu}\frac{\partial}{\partial\theta}-g_{_X}\frac{\partial x^\nu}{\partial x'^\mu}\left(X_\nu-\frac{1}{g_{_X}}\partial_\nu\alpha(x)\right)\left(\frac{\partial x^\mu}{\partial\theta'}\frac{\partial}{\partial x^\mu}+\frac{\partial\theta}{\partial\theta'}\frac{\partial}{\partial\theta}\right)\nonumber\\
&=&\frac{\partial x^\nu}{\partial x'^\mu}\hat{\partial}_\nu,\nonumber\\
\partial_\theta\longrightarrow\partial'_\theta&=&\frac{\partial x^\mu}{\partial\theta'}\frac{\partial}{\partial x^\mu}+\frac{\partial\theta}{\partial\theta'}\frac{\partial}{\partial\theta},\nonumber\\
&=&\frac{\partial\theta}{\partial\theta'}\partial_\theta=\partial_\theta.
\end{eqnarray}
It should be noted that $X_\mu$ would disappear if the fiber bundle spacetime $M_5$ is trivial or $M_5$ is globally a direct product $M_4\times\mathrm{U}(1)$ because $T_pM_5=T_xM_4\oplus T_g\mathrm{U}(1)$ with $x\in M_4$ and $g\in\mathrm{U}(1)$. Similarly, the cotangent space $T^*_pM_5$ which is dual to $T_pM_5$ is always decomposed into a direct sum of horizontal cotangent subspace $H^*_pM_5$ and vertical cotangent subspace $V^*_pM_5$ as
\begin{equation}
T^*_pM_5=V^*_pM_5\oplus H^*_pM_5,
\end{equation}
without reference to the local coordinate system. Note that, $H^*_pM_5$ and $V^*_pM_5$ are dual to $H_pM_5$ and $V_pM_5$, respectively. The covariant bases, which span the cotangent subspaces $H^*_pM_5$ and $V^*_pM_5$, are given by $\{dx^\mu\}$ and $\{d\theta+g_{_X}X_\mu dx^\mu\}$, respectively, which transform under the general coordinate transformation (\ref{ba-gct}) as
\begin{eqnarray}
dx^\mu&\longrightarrow& dx'^\mu=\frac{\partial x'^\mu}{\partial x^\nu}dx^\nu,\nonumber\\
d\theta+g_{_X}X_\mu dx^\mu&\longrightarrow& d\theta'+g_{_X}X'_\mu dx'^\mu=\frac{\partial\theta'}{\partial\theta}(d\theta+g_{_X}X_\mu dx^\mu)=d\theta+g_{_X}X_\mu dx^\mu.
\end{eqnarray}
Because of the above natural decomposition, an inner product $G$ on the spacetime $M_5$ is defined as
\begin{equation}\label{bulk-mertic}
  G(V_1,V_2)=G_H(V_{1H},V_{2H})+G_V(V_{1V},V_{2V}),
\end{equation}
where $V_{1H}$ ($V_{2H})$ and $V_{1V}$ ($V_{2V}$) are the horizon and vertical components of the vector $V_1$ ($V_2$), respectively. In the expression (\ref{bulk-mertic}), $G_H$ is called the horizontal metric which is a tensor field belonging the space $H^*_pM_5\otimes H^*_pM_5$ and given by
\begin{equation}
  G_H=g_{\mu\nu}dx^\mu dx^\nu,
\end{equation}
where $g_{\mu\nu}$ transforms under the general coordinate transformation (\ref{ba-gct}) as
\begin{eqnarray}
g_{\mu\nu}\longrightarrow\frac{\partial x^\rho}{\partial x'^\mu}\frac{\partial x^\lambda}{\partial x'^\nu}g_{\rho\lambda}.
\end{eqnarray}
Whereas, $G_V$ is called the vertical metric which is a tensor field belonging the space $V^*_pM_5\otimes V^*_pM_5$ and given by
\begin{equation}
G_V=-T^2\frac{(d\theta+g_{_X}X_\mu
dx^\mu)^2}{\Lambda^2},
\end{equation}
where the field $T$ is related to the geometric size of the fiber and $\Lambda$ is a constant of the energy dimension. Under the general coordinate transformation (\ref{ba-gct}), the field $T$ transforms as
\begin{equation}
T\longrightarrow\frac{\partial\theta}{\partial\theta'}T=T.
\end{equation}
With the horizontal metric $G_H$ written in the local flat form as
\begin{eqnarray}
G_H=\eta_{\alpha\beta}\hat{e}^\alpha\hat{e}^\beta,
\end{eqnarray}
where $\hat{e}^\alpha={e^\alpha}_\mu dx^\mu$ which are dual to $\hat{e}_\alpha={e_\alpha}^\mu\hat{\partial}_\mu$ with ${e_\alpha}^\mu$ called the vierbeins, we can see that the spacetime metric is invariant under the following change 
\begin{eqnarray}
\hat{e}^\alpha\longrightarrow{\Lambda^\alpha}_\beta\hat{e}^\beta,
\end{eqnarray}
where ${\Lambda^\alpha}_\beta\in SO(1,3)$. In this way, the right-handed and left-handed Weyl spinor fields may be defined in the fiber bundle spacetime $M_5$, which does not happen to a general five-dimensional manifold. In the base $\left\{\hat{\partial}_\mu,\partial_\theta\right\}\equiv\left\{\partial_M\right\}$, the coefficients of the Christoffel connection and the Riemann curvature tensor are given by
\begin{eqnarray}
\Gamma^P_{MN}&=&\frac{G^{PQ}}{2}\left(\partial_MG_{NQ}+\partial_NG_{MQ}-\partial_QG_{MN}\right)
+\frac{G^{PQ}}{2}\left(C^O_{QM}G_{ON}+C^O_{QN}G_{OM}\right)+\frac{C^P_{MN}}{2},
\nonumber \\
\mathcal{R}^O_{MPN}&=&\partial_N[\Gamma^O_{PM}]-\partial_P[\Gamma^O_{NM}]+
\Gamma^Q_{PM}\Gamma^O_{NQ}-\Gamma^Q_{NM}\Gamma^O_{PQ}+C^Q_{PN}\Gamma^O_{QM},
\end{eqnarray}
where $C^P_{MN}$ are the non-holonomic functions determining the commutator of any two frame fields as
\begin{equation}
\left[\partial_M,\partial_N\right]=C^P_{MN}\partial_P,
\end{equation}
and
\begin{equation}
  G_{MN}=\textrm{diag}\left(g_{\mu\nu},-\frac{T^2}{\Lambda^2}\right),\ \ G^{MN}=\textrm{diag}\left(g^{\mu\nu},-\frac{\Lambda^2}{T^2}\right).
\end{equation}

Let now us introduce the action for the fields which describe the spacetime $M_5$ as 
\begin{eqnarray}
S_{\text{spacetime}}&=&M^3_*\int d^4xd\theta|\textrm{det}G|^{1/2}
\left[\frac{\mathcal{R}}{2}+G^{\theta\theta}\left(\partial_\theta T\right)\left(\partial_\theta T\right)-f(T)\right]\nonumber\\
&=&\frac{M^3_*}{\Lambda}\int
d^4xd\theta\left[-\frac{g^2_{_X}T^3}{8\Lambda^2}X_{\mu\nu}X^{\mu\nu}+\frac{1}{T}\left(\hat{\partial}_\mu T\right)\left(\hat{\partial}^\mu T\right)-\frac{1}{T}\left(\hat{\partial}_\theta T\right)\left(\hat{\partial}_\theta T\right)-V(T)\right],\nonumber\\\label{EH-act}
\end{eqnarray}
where $M_*$ is the five-dimensional Planck scale, $\mathcal{R}$ is the scalar curvature of the spacetime $M_5$, $\hat{\partial}_\theta\equiv\Lambda\partial_\theta$, $X_{\mu\nu}=\partial_\mu X_\nu-\partial_\nu X_\mu$ is the field strength tensor of the gauge field $X_\mu$, and $V(T)\equiv Tf(T)$ is the potential of the field $T$ which can be constructed at the tree level because $T$ transforms as a scalar under the general coordinate transformation (\ref{ba-gct}) and is generally given by
\begin{eqnarray}
V(T)=\Lambda^2T\left[a_0+a_1T+a_2T^2+a_3T^3+a_4T^4+\cdots\right],
\end{eqnarray}
where $a_{i}$ are the dimensionless constants. Note that, in the action (\ref{EH-act}) we have ignored the four-dimensional gravitation because it does not play the role in the present work and hence $g_{\mu\nu}=\eta_{\mu\nu}$, and the second term in the first line of (\ref{EH-act}) (which does not appear in $\mathcal{R}$) represents the kinetic term of the field $T$ corresponding to the vertical direction in the spacetime $M_5$. With the change of the field variable as
\begin{eqnarray}
T'=2\left(\frac{2M^3_*T}{\Lambda}\right)^{1/2},
\end{eqnarray}
one can rewrite the action (\ref{EH-act}) as
\begin{eqnarray}\label{bulk-act}
S_{\text{spacetime}}&=&\int
d^4xd\theta\left[-\frac{g^2_{_X}T'^{6}}{2^{12}M^6_*}X_{\mu\nu}X^{\mu\nu}+
\frac{1}{2}\left(\hat{\partial}_\mu T'\right)\left(\hat{\partial}^\mu T'\right)-\frac{1}{2}\left(\hat{\partial}_\theta T'\right)\left(\hat{\partial}_\theta T'\right)-V(T')\right],\nonumber\\
\end{eqnarray}
where
\begin{eqnarray}
V(T')\simeq\mu^2_{T'}T'^2+\lambda_{T'}T'^4,\label{sta-pot}
\end{eqnarray}
with $\mu^2_{T'}\equiv a_0\Lambda^2/8$, $\lambda_{T'}=a_1\Lambda^3/64M^3_*$, and the higher order terms in $V(T')$ suppressed by $1/M_*$. The potential $V(T')$ allows the stabilization of the field $T'$ through
which the size of the fifth dimension is physically fixed at even the classical level. The nonzero vacuum expectation value (VEV) of the field $T'$ is determined by
\begin{eqnarray}
\langle T'\rangle\simeq\sqrt{-\frac{\mu^2_{T'}}{2\lambda_{T'}}}.
\end{eqnarray}
The fluctuation around this vaucuum is described by
\begin{eqnarray}
T'=\frac{1}{\sqrt{2}}\left[\langle T'\rangle+r(x,\theta)\right],
\end{eqnarray}
where the fluctuation field $r(x,\theta)$ has the mass, $m^2_r=-\mu^2_{T'}/2$. The fluctuation field $r(x,\theta)$ is heavy enough to decouple at the low energy region. Hence, we may consider theory at the vacuum $\langle T'\rangle$ corresponding to the geometric size of the fiber determined by the radius $R=\langle T'\rangle^2/8M^3_*$. 

In this way, the first term in (\ref{bulk-act}) becomes
\begin{eqnarray}
S_X=\frac{\pi g^2_{_X}\langle T'\rangle^{6}}{2^{9}M^6_*}\int
d^4x\left(-\frac{1}{4}X_{\mu\nu}X^{\mu\nu}\right).
\end{eqnarray}
To get the canonically normalized action for the gauge field $X_\mu$, the coupling constant
$g_{_X}$ should be determined as
\begin{equation}
g_{_X}=\sqrt{\frac{1}{2\pi}}\frac{2^5M^3_*}{\langle T'\rangle^3}=\frac{\sqrt{2}}{M_{\text{Pl}}R},\label{gX-coupling}
\end{equation}
where $M_{\text{Pl}}$ is the reduced four-dimensional Planck scale related to $M_*$ and $R$ as, $M^2_{\text{Pl}}=2\pi RM^3_*$. This implies that the interacting strength of the gauge field $X_\mu$ to other fields should be very small if  the inverse bulk radius is much smaller than the reduced four-dimensional Planck scale, $R^{-1}\ll M_{\text{Pl}}$. 

\subsection{Realistic model}
The model which will study is a set of the fields propagating dynamically in the spacetime $M_5$ and respecting for the SM gauge symmetry group. The fermion content is given as
\begin{eqnarray}
L_{a}(x,e^{i\theta})&=&\frac{1}{\sqrt{2\pi R}}\left(%
\begin{array}{c}
  \nu_{{aL}}(x) \\
  e_{aL}(x) \\
\end{array}%
\right)e^{iX_{L_a}\theta}\equiv\frac{L_a(x)}{\sqrt{2\pi R}}e^{iX_{L_a}\theta}\sim \left(1,2,-\frac{1}{2}\right),\nonumber
\end{eqnarray}
\begin{eqnarray}
E_{aR}(x,e^{i\theta})&=&\frac{e_{aR}(x)}{\sqrt{2\pi R}}e^{iX_{E_a}\theta}\sim \left(1,1,-1\right),\nonumber
\end{eqnarray}
\begin{eqnarray}
N_{aR}(x,e^{i\theta})&\sim& \left(1,1,0\right),\nonumber
\end{eqnarray}
\begin{eqnarray}
Q_{a}(x,e^{i\theta})&=&\frac{1}{\sqrt{2\pi R}}\left(%
\begin{array}{c}
  u_{aL}(x) \\
  d_{aL}(x) \\
\end{array}%
\right)e^{iX_{Q_a}\theta}\equiv\frac{Q_a(x)}{\sqrt{2\pi R}}e^{iX_{Q_a}\theta}\sim \left(3,2,\frac{1}{6}\right),\nonumber
\end{eqnarray}
\begin{eqnarray}
D_{aR}(x,e^{i\theta})&=&\frac{d_{aR}(x)}{\sqrt{2\pi R}}e^{iX_{D_a}\theta}\sim \left(3,1,-\frac{1}{3}\right),\nonumber
\end{eqnarray}
\begin{eqnarray}
U_{aR}(x,e^{i\theta})=\frac{u_{aR}(x)}{\sqrt{2\pi R}}e^{iX_{U_a}\theta}\sim \left(3,1,\frac{2}{3}\right),\label{the-dep}
\end{eqnarray}
where the numbers given in parentheses are the quantum numbers corresponding to the gauge symmetries $\{\mathrm{SU}(3)_C$, $\mathrm{SU}(2)_L$, $\mathrm{U}(1)_Y\}$, respectively, and $a=1,2,3$ are the generation indices. As indicated in Ref. \cite{Nam2019a}, because the fermion fields except the right-handed neutrinos $N_{aR}$ transform non-trivially under the SM gauge symmetry group, the vertical kinetic term describing their propagation along the vertical direction in the spacetime $M_5$ is not invariant and thus is forbidden. As a result, these fermion fields themselves have no the terms determining the $\theta$-dependence or the dynamics along the vertical direction in the spacetime $M_5$. And, thus their $\theta$-dependence must be determined by a certain special property which nothing but they are invariant under the active action of the Lie group $\textrm{U}(1)$. In (\ref{the-dep}), the numbers ($X_{L_a}$, $X_{E_a}$, $X_{Q_a}$, $X_{D_a}$, $X_{U_a}$) are quantum numbers characterizing the active action of the Lie group $\textrm{U}(1)$ on the corresponding fermion fields, and the fields [$L_a(x)$, $e_{aR}(x)$, $Q_a(x)$, $d_{aR}(x)$, $u_{aR}(x)$] should be identified as the SM fermion fields. 

It is easily to see from (\ref{the-dep}) that the transforming parameters of the SM gauge symmetry group  are completely independent on the fiber coordinate $\theta$ but only dependent on the $x$-coordinates. Since it leads to the simplest form for the gauge fields of the SM gauge symmetry group as
\begin{eqnarray}
  G_{aM} &=& \left(\frac{G_{a\mu}(x)}{\sqrt{2\pi R}},0\right),\nonumber\\
  W_{iM} &=& \left(\frac{W_{i\mu}(x)}{\sqrt{2\pi R}},0\right),\nonumber\\
  B_M &=& \left(\frac{B_\mu(x)}{\sqrt{2\pi R}},0\right).
\end{eqnarray} 
Bulk action for the gauge boson and fermion fields, up to the gauge fixing and ghost terms, is given by
\begin{eqnarray}
S^{\text{bulk}}_{\text{FG}}&=&\int dx^4d\theta\sqrt{|\textrm{det}G|}\left(\mathcal{L}^{\text{bulk}}_{\text{gauge}}+\mathcal{L}^{\text{bulk}}_{\text{fer}}\right),\nonumber
\end{eqnarray}
\begin{eqnarray}
\mathcal{L}^{\text{bulk}}_{\text{gauge}}&=&-\frac{1}{4}G_{aMN}G^{MN}_a-\frac{1}{4}W_{iMN}W^{MN}_i-\frac{1}{4}B_{MN}B^{MN}+\frac{M^3_*}{2}\mathcal{R}\Big|_{\langle T'\rangle},\nonumber\label{gaug-term}\\
\mathcal{L}^{\text{bulk}}_{\text{fer}}&=&\sum_F\bar{F}i\gamma^\mu\hat{D}_\mu F+\bar{N}_{aR}i\gamma^\mu\hat{\partial}_\mu N_{aR}+\frac{1}{2\Lambda}\left(\partial^\theta\bar{N}^C_{aR}\partial_\theta N_{aR}-M^2_{N_a}
\bar{N}^C_{aR}N_{aR}+\textrm{H.c.}\right),\label{SMfers}
\end{eqnarray}
where $\{G_{aMN}, W_{iMN}, B_{MN}\}$ are the field strength tensors of the gauge fields $\{G_{aM}, W_{iM}, B_M\}$, which have the non-zero components given by (up to a normalized factor)
\begin{eqnarray}
  G_{a\mu\nu} &=& \partial_\mu G_{a\nu}-\partial_\nu G_{a\mu}+g_sf_{abc}A_{b\mu}A_{c\nu},\nonumber\\
  W_{i\mu\nu} &=& \partial_\mu W_{i\nu}-\partial_\nu W_{i\mu}+g\varepsilon_{ijk}W_{j\mu}W_{k\nu},\nonumber\\
  B_{\mu\nu} &=& \partial_\mu B_\nu-\partial_\nu B_\mu,
\end{eqnarray}
$M_{N_a}$ are the vertical mass parameters of the right-handed neutrinos $N_{aR}$ which are naturally in the order of the scale $\Lambda$, and the covariant derivative $\hat{D}_\mu$ reads
\begin{equation}
\hat{D}_\mu=\hat{\partial}_\mu-ig_s\frac{\lambda^a}{2}G_{a\mu}-ig\frac{\sigma^i}{2}W_{i\mu}-ig'Y_FB_\mu,
\end{equation}
with $\hat{\partial}_\mu\equiv\partial_\mu-g_{_X}X_\mu\partial_\theta$ and $\{g_s,g,g'\}$ to be coupling constants of the gauge symmetries $\{\mathrm{SU}(3)_C, \mathrm{SU}(2)_L, \mathrm{U}(1)_Y\}$. Note that, the sum in $\mathcal{L}^{\text{bulk}}_{\text{fer}}$ is taken over all fermion fields, except the right-handed neutrinos $N_{aR}$, given in (\ref{the-dep}). From the bulk action (\ref{SMfers}), we can find the effective action in the four-dimensional effective spacetime as
\begin{eqnarray}
S^{\text{eff}}_{\text{FG}}=\int dx^4\left(-\frac{1}{4}G_{a\mu\nu}G^{a\mu\nu}-\frac{1}{4}W_{i\mu\nu}W^{i\mu\nu}-\frac{1}{4}B_{\mu\nu}B^{\mu\nu}-\frac{1}{4}X_{\mu\nu}X^{\mu\nu}+\sum_f\bar{f}i\gamma^\mu D_\mu f+\mathcal{L}_N\right),\nonumber
\end{eqnarray}
\begin{eqnarray}
\mathcal{L}_N&=&\mathcal{L}_\nu+\mathcal{L}_\psi+\mathcal{L}_\chi+\mathcal{L}_{\text{int}},\nonumber\\
\mathcal{L}_\nu&=&\bar{\nu}_{aR}
i\gamma^\mu\partial_\mu\nu_{aR}-\frac{M_{a0}}{2}\bar{\nu}^C_{aR}\nu_{aR}+\textrm{H.c.},\nonumber\\
\mathcal{L}_\psi&=&\sum_{n=1}^{\infty}\left(\bar{\psi}_{naR}
i\gamma^\mu\partial_\mu\psi_{naR}-\frac{M_{an}}{2}\bar{\psi}^C_{naR}\psi_{naR}+\textrm{H.c.}\right),\nonumber\\
\mathcal{L}_\chi&=&\sum_{n=1}^{\infty}\left(\bar{\chi}_{naR}
i\gamma^\mu\partial_\mu\chi_{naR}-\frac{M_{an}}{2}\bar{\chi}^C_{naR}\chi_{naR}+\textrm{H.c.}\right),\nonumber\\
\mathcal{L}_{\text{int}}&=&ig_{_X}\sum_{n=1}^{\infty}n\Big(\bar{\chi}_{naR}\gamma^\mu\psi_{naR}-\bar{\psi}_{naR}\gamma^\mu\chi_{naR}\Big)X_\mu,\label{eff-act}
\end{eqnarray}
where the four-dimensional effective fermion $f(x)$ is related to the five-dimensional fundamental fermion $F(x,e^{i\theta})$ as
\begin{eqnarray}
F(x,e^{i\theta})&=&\frac{f(x)}{\sqrt{2\pi R}}e^{iX_F\theta},
\end{eqnarray}
the covariant derivative $D_\mu$ reads
\begin{equation}\label{eff-covder}
D_\mu=\partial_\mu-ig_s\frac{\lambda^a}{2}G_{a\mu}-ig\frac{\sigma^i}{2}W_{i\mu}-ig'Y_fB_\mu-ig_{_X}X_fX_\mu,
\end{equation}
(with $Y_F$ and $X_F$ to be replaced by $Y_f$ and $X_f$, respectively, for a convenient reason), $\nu_{aR}(x)$ are identified as the usual right-handed neutrinos and $\{\psi_{naR},\chi_{naR}\}$ are their Kaluza-Klein (KK) excitations whose masses are given by
\begin{eqnarray}
  M_{a0} &=& \frac{M^2_{N_a}}{\Lambda}\sim\Lambda,\nonumber\\
  M_{an} &=& \frac{1}{\Lambda}\left(M^2_{N_a}+\frac{n^2}{R^2}\right)\sim\Lambda.
\end{eqnarray}

The effective action (\ref{eff-act}) looks like an extension of the SM based on the gauge symmetry group $\mathrm{SU}(3)_C\otimes\mathrm{SU}(2)_L\otimes\mathrm{U}(1)_Y\otimes\mathrm{U}(1)_X$. Of course, we know here that $\mathrm{U}(1)_X$ is not the fundamental gauge symmetry but it is emerged from the short-distance structure of the spacetime. The emergent $\mathrm{U}(1)_X$ charges of the SM fermions are defined in Table \ref{Xcharge-number}.
\begin{table}[!h]
\begin{center}
\begin{tabular}{|c|c|c|c|c|c|c|c|}
\hline
Fermion $f$ & $\nu_{aL}$ & $e_{aL}$ & $e_{aR}$ & $u_{aL}$ & $d_{aL}$ & $u_{aR}$ & $d_{aR}$ \\
\hline
$X_f$ & $X_{L_a}$ & $X_{L_a}$ & $X_{E_{a}}$ & $X_{Q_a}$ & $X_{Q_a}$ & $X_{U_a}$ & $X_{D_a}$ \\
\hline
\end{tabular}
\caption{The $\mathrm{U}(1)_X$ charges of the SM fermions.}\label{Xcharge-number}
\end{center}
\end{table}
The coupling constant $g_{_X}$ corresponding to the emergent $\mathrm{U}(1)_X$ gauge group is determined in terms of the radius $R$ of the fiber and the reduced four-dimensional Planck scale $M_{\text{Pl}}$ as given in (\ref{gX-coupling}). This relation suggests that the coupling constant $g_{_X}$ is completely fixed by the difference between the mass of the zero mode and that of its first KK excitation. This is one of the essential predictions in this emergent $\mathrm{U}(1)_X$ model.

The scalar sector is given by
\begin{eqnarray}
 H(x,e^{i\theta})&=&\frac{1}{\sqrt{2\pi R}}\left(%
\begin{array}{c}
  \phi^+(x) \\
  \phi^0(x) \\
\end{array}%
\right)e^{iX_H\theta}\equiv\frac{\phi(x)}{\sqrt{2\pi R}}e^{iX_H\theta}\sim \left(1,2,\frac{1}{2}\right),\nonumber\\
\Phi(x,e^{i\theta})&=&\frac{\varphi(x)}{\sqrt{2\pi R}}e^{iX_\Phi\theta}\sim \left(1,1,0\right).\label{scalar-sector}
\end{eqnarray}
It is important to note that comparing to Ref. \cite{Nam2019a} we have modified the scalar doublet $H$ in such a way that $H$ is invariant under the active action of the Lie group $\textrm{U}(1)$ and thus has the specific $\theta$-dependence given as in (\ref{scalar-sector}). As seen later, this modification leads to no the couplings between the new gauge boson and all neutrinos, which allows to avoid the stringent constraints from the $\nu_e-e^-$ scattering experiments such as TEXONO experiment. The bulk action for the scalar sector is given by
\begin{equation}
S[H,\Phi]=\int
d^4xd\theta\sqrt{|\textrm{det}G|}\left[\Big|\Big(\hat{\partial}_\mu-ig\frac{\sigma^i}{2}W_{i\mu}-i\frac{g'}{2}B_\mu\Big)H\Big|^2+\left|\hat{\partial}_\mu\Phi\right|^2-V(H,\Phi)\right],\label{H-Phi-act}
\end{equation}
where the scalar potential is given by
\begin{equation}
  V(H,\Phi)=\mu^2_1H^\dagger H+\bar{\lambda}_1(H^\dagger H)^2+\mu^2_2\Phi^\dagger\Phi+\bar{\lambda}_2(\Phi^\dagger\Phi)^2+\bar{\lambda}_3(H^\dagger H)(\Phi^\dagger\Phi).
\end{equation}
For simplicity, we set $X_\Phi=1$ in this work. (If $X_\Phi\neq1$, the related constraints will be imposed on the coupling constant by scaling as $g_{_X}/X_\Phi$.) The effective action $S_{\text{eff}}$ in the four-dimensional effective spacetime reads
\begin{eqnarray}\label{H-Phi-act}
S_{\text{eff}}&=&\int
d^4x\left[\left|D_\mu\phi\right|^2+\left|(\partial_\mu-ig_{_X}X_\mu)\varphi\right|^2-V(\phi,\varphi)\right],\nonumber\\
 V(\phi,\varphi)&=&\mu^2_1\phi^\dagger\phi+\lambda_1(\phi^\dagger\phi)^2+\mu^2_2\varphi^\dagger\varphi+\lambda_2(\varphi^\dagger\varphi)^2+\lambda_3(\phi^\dagger\phi)(\varphi^\dagger\varphi).
\end{eqnarray}
where $D_\mu=\partial_\mu-ig\frac{\sigma^i}{2}W_{i\mu}-i\frac{g'}{2}B_\mu-ig_{_X}X_H X_\mu$, $\lambda_1=\bar{\lambda}_1/2\pi R$, $\lambda_2=\bar{\lambda}_2/2\pi R$, and $\lambda_3=\bar{\lambda}_3/2\pi R$.

The gauge symmetry $\mathrm{SU}(2)_L\otimes\mathrm{U}(1)_Y$ is spontaneously broken due to that the scalar doublet $\phi$ develops the VEV, whereas the emergent $\mathrm{U}(1)_X$ gauge symmetry is spontaneously broken by the the VEV of the scalar $\varphi$. These VEVs are given by
\begin{eqnarray}
  \langle\phi\rangle = \frac{1}{\sqrt{2}}\left(
                                             \begin{array}{c}
                                               0 \\
                                               v \\
                                             \end{array}
                                           \right),\ \ \ \  \langle\varphi\rangle=\frac{v'}{\sqrt{2}},
  \end{eqnarray}
where
\begin{eqnarray}
 v^2=2\frac{2\lambda_2\mu^2_1-\lambda_3\mu^2_2}{\lambda^2_3-4\lambda_1\lambda_2},\ \ \ \ v'^2=2\frac{2\lambda_1\mu^2_2-\lambda_3\mu^2_1}{\lambda^2_3-4\lambda_1\lambda_2}.
\end{eqnarray}
We expand these scalar fields around the vacuum as
\begin{eqnarray}
\phi = \left(
\begin{array}{c}
w^+(x) \\
\frac{v+h(x)+iz(x)}{\sqrt{2}} \\
\end{array}
\right),\ \ \ \ \varphi=\frac{v'+h'(x)+iz'(x)}{\sqrt{2}}.
\end{eqnarray}
Here, the $CP$-odd fields $w^+(x)$, $z(x)$ and $z'(x)$ are Nambu-Goldstone bosons which should be absorbed
by the weak gauge bosons and the $\mathrm{U}(1)_X$ gauge boson. The $CP$-even fields mixes together where
their squared mass matrix is given by
\begin{eqnarray}
 \mathcal{L}_{\text{mass}}(h,h')=\frac{1}{2}\left(
                                      \begin{array}{cc}
                                        h & h' \\
                                      \end{array}
                                    \right)\left(
                                             \begin{array}{cc}
                                               2\lambda_1v^2 & \lambda_3vv' \\
                                               \lambda_3vv' & 2\lambda_2v'^2 \\
                                             \end{array}
                                           \right)\left(
                                                    \begin{array}{c}
                                                      h \\
                                                      h' \\
                                                    \end{array}
                                                  \right).
\end{eqnarray}
The physical states are found as, $h_1=c_\alpha h-s_\alpha h'$ and $h_2=s_\alpha h+c_\alpha h'$, corresponding to the following masses
\begin{equation}
  m^2_{h_1,h_2}=\lambda_1v^2+\lambda_2v'^2\mp\left[\left(\lambda_1v^2-\lambda_2v'^2\right)^2+\lambda^2_3v^2v'^2\right]^{1/2}.
\end{equation}
The mixing angle $\alpha$ is defined as, $\sin(2\alpha)=2\lambda_3vv'/(m^2_{h_2}-m^2_{h_1})$. It is constrained by the measurements of the Higgs production cross section and its decay branching ratio at the LHC as $s_\alpha\lesssim0.2$ \cite{Tanabashi2018,Nomura2018} which leads to the following constraint
\begin{eqnarray}
\left|\frac{\lambda_3vv'}{\lambda_1v^2-\lambda_2v'^2}\right|\lesssim0.426.\label{lam3-constr}
\end{eqnarray}

The mixing mass matrix between the bosons $W^3_\mu$, $B_\mu$ and $X_\mu$ is given by
\begin{eqnarray}
M^2=\frac{1}{4}\left(%
\begin{array}{ccc}
  g^2v^2 & -gg'v^2 & -2gg_{_X}X_Hv^2 \\
  -gg'v^2 & g'^2v^2 & 2g'g_{_X}X_Hv^2 \\
  -2gg_{_X}X_Hv^2 & 2g'g_{_X}X_Hv^2 & 4g^2_{_X}(X^2_Hv^2+v'^2) \\
\end{array}%
\right),
\end{eqnarray}
This mass matrix is diagonalized by two matrices $V$ and $U$ as
\begin{eqnarray}
\text{Diag}\left(M^2_Z, 0, M^2_{Z'}\right)=U^TV^TM^2VU,
\end{eqnarray}
where
\begin{eqnarray}
M^2_{Z,Z'}&=&\frac{1}{8}\left[\left(g^2+g'^2\right)v^2+4g^2_{_X}\left(v'^2+X^2_Hv^2\right)\right.\nonumber\\
&&\left.\pm\sqrt{\left[\left(g^2+g'^2\right)v^2+4g^2_{_X}\left(v'^2+X^2_Hv^2\right)\right]^2-16\left(g^2+g'^2\right)g^2_{_X}v^2v'^2}\right],\label{neut-bos-mass}
\end{eqnarray}
corresponding to the physical states which are the SM neutral gauge boson $Z$ and the new gauge boson $Z'$, respectively, and the matrices $V$ and $U$ are given by
\begin{eqnarray}
V=\left(%
\begin{array}{ccc}
  c_W & s_W & 0 \\
  -s_W & c_W & 0 \\
  0 & 0 & 1 \\
\end{array}%
\right),\ \ \ \ U=\left(%
\begin{array}{ccc}
  c_\beta & 0 & s_\beta \\
  0 & 1 & 0 \\
  -s_\beta & 0 & c_\beta \\
\end{array}%
\right),
\end{eqnarray}
where the mixing angle $\beta$ is determined by
\begin{eqnarray}
\tan(2\beta)=\frac{4g_{_X}X_H\sqrt{g^2+g'^2}}{(g^2+g'^2)-4g^2_X(X^2_H+v'^2/v^2)}.
\end{eqnarray}
Because $g_{_X}$ is very small to explain the $^8$Be anomaly, thus the mixing angle $\beta$ should be small which is approximately given by
\begin{equation}
\beta\simeq\frac{2g_{_X}X_H}{\sqrt{g^2+g'^2}}.
\end{equation}

The new gauge boson $Z'$ couples to the SM fermions through the neutral current as
\begin{equation}
\mathcal{L}\supset\sum_{f}\bar{f}\gamma^\mu\left(C_{f,V}+C_{f,A}\gamma^5\right)fZ'_\mu,
\end{equation}
where the vector and axial couplings are given by
\begin{eqnarray}
C_{f,V}&=&\frac{gs_\beta}{2c_W}\left(T^3_{f_L}-2s^2_WQ_f\right)+c_\beta g_{_X}\frac{X_{f_R}+X_{f_L}}{2},\nonumber\\
&\simeq&g_{_X}\left[X_H\left(T^3_{f_L}-2s^2_WQ_f\right)+\frac{X_{f_R}+X_{f_L}}{2}\right],\nonumber\\
C_{f,A}&=&-\frac{gs_\beta}{2c_W}T^3_{f_L}+c_\beta g_{_X}\frac{X_{f_R}-X_{f_L}}{2},\nonumber\\
&\simeq&-g_{_X}\left(X_HT^3_{f_L}+\frac{X_{f_L}-X_{f_R}}{2}\right),\label{V-A-coup}
\end{eqnarray}
where $Q_f$ and $T^3_{f_L}$ refer to the electric charge and the weak isospin of the fermion $f$, respectively. For the neutrinos, we have $C_{\nu_a,V}=C_{\nu_a,A}=g_{_X}\left(X_H+X_{L_a}\right)/2$. We expect that there are no the $Z'$ couplings to all neutrinos. This is satisfied if $X_H=-X_{L_a}$ and $X_{L_1}=X_{L_2}=X_{L_3}$.

With the vector and axial couplings given in (\ref{V-A-coup}) and the $\mathrm{U}(1)_X$ charges of the SM fermions given in Table 1 of Ref. \cite{Nam2019a}, one can find that $C_{e,A}=C_{\mu,A}=C_{\tau,A}=0$ which allows to evade the very stringent constraints such as the atomic parity violation in Cesium or the $(g-2)_{e,\mu}$ constraints. Whereas, in analogy the axial couplings for the quarks vanish all. However, a model with primarily vector couplings to the quarks should have the relatively large $g_{_X}$ coupling constant to explain the $^8$Be anomaly, which is difficult to evade various experimental bounds. Thus, we will modify the $\mathrm{U}(1)_X$ charges of the quarks to obtain a consistent model to explain the $^8$Be anomaly. Because the $\mathrm{U}(1)_X$ charges of the SM fermions given in Table 1 of Ref. \cite{Nam2019a} is the unique solution with the universality among the generations, modifying the $\mathrm{U}(1)_X$ charges of the quarks suggests that there is no the universality of the $\mathrm{U}(1)_X$ charge among the generations of the quarks. The absence of the nontrivial anomalies associated with $\mathrm{U}(1)_X$ lead to the following equations for the $\mathrm{U}(1)_X$ charges of the SM fermions
\begin{eqnarray}
\sum^3_{a=1}\left(2X_{Q_a}-X_{u_{aR}}-X_{d_{aR}}\right)&=&0,\nonumber\\
\sum^3_{a=1}X_{Q_a}+X_L&=&0,\nonumber\\
\sum^3_{a=1}\left(X_{Q_a}-8X_{u_{aR}}-2X_{d_{aR}}\right)+9\left(X_L-2X_e\right)&=&0,\nonumber\\
\sum^3_{a=1}\left(X^2_{Q_a}-2X^2_{u_{aR}}+X^2_{d_{aR}}\right)-3\left(X^2_L-X^2_e\right)&=&0,\nonumber\\
\sum^3_{a=1}\left(2X^3_{Q_a}-X^3_{u_{aR}}-X^3_{d_{aR}}\right)+\left(2X^3_L-X^3_e\right)&=&0,\nonumber\\
\sum^3_{a=1}\left(2X_{Q_a}-X_{u_{aR}}-X_{d_{aR}}\right)+\left(2X_L-X_e\right)&=&0,
\end{eqnarray}
where
\begin{eqnarray}
X_{L_1}&=&X_{L_2}=X_{L_3}\equiv X_L,\nonumber\\
X_{e_{1R}}&=&X_{e_{2R}}=X_{e_{3R}}\equiv X_e.
\end{eqnarray}
It should be noted here that, the usual right-handed neutrinos $\nu_{aR}$ do not contribute to the above six anomalies because they have no the charges under both the SM gauge symmetry group and the $\mathrm{U}(1)_X$ one. In addition, their KK counterparts $\{\psi_{naR},\chi_{naR}\}$ do not contribute to the first four anomalies because they have no the charges under the SM gauge symmetry group. Whereas, because $\{\psi_{naR},\chi_{naR}\}$ couple to the gauge boson $X_\mu$ in a unusual way given in (\ref{eff-act}), they should not appear in the last two anomalies. One solution of this system of equations is found as
\begin{eqnarray}\label{charge}
X_{Q_1}&=&\frac{X_{Q_2}}{4}=-\frac{X_{Q_3}}{2}=\frac{y}{3},\nonumber\\
X_{u_{1R}}&=&\frac{X_{u_{2R}}}{4}=\frac{X_{u_{3R}}}{7}=\frac{y}{3},\nonumber\\
X_{d_{1R}}&=&-\frac{X_{d_{2R}}}{2}=-\frac{X_{d_{3R}}}{5}=\frac{y}{3},\nonumber\\
X_{L_1}&=&X_{L_2}=X_{L_3}=-y,\nonumber\\
X_{e_{1R}}&=&X_{e_{2R}}=X_{e_{3R}}=-2y,
\end{eqnarray}
where $y$ is a free parameter. 

Note that, the mass eigenstates $u'_{L,R}=(u,c,t)^T_{L,R}$ and $d'_{L,R}=(d,s,b)^T_{L,R}$ are related to the  weak states $u_{L,R}=(u_1,u_2,u_3)^T_{L,R}$ and $d_{L,R}=(d_1,d_2,d_3)^T_{L,R}$ as
\begin{eqnarray}
u'_L=\mathcal{U}^\dagger_u u_L,\ \ \ \  u'_R=\mathcal{V}^\dagger_u u_R\ \ \ \ d'_L=\mathcal{U}^\dagger_d d_L,\ \ \ \  d'_R=\mathcal{V}^\dagger_d d_R.
\end{eqnarray}
In the basis of the mass eigenstates, the couplings of the gauge boson $Z'$ with the quarks are given by
\begin{eqnarray}
\mathcal{L}\supset\left(\bar{u'}_L\gamma^\mu\Gamma_{L,u}u'_L+\bar{u'}_R\gamma^\mu\Gamma_{R,u}u'_R+\bar{d'}_L\gamma^\mu\Gamma_{L,d}d'_L+\bar{d'}_R\gamma^\mu\Gamma_{R,d}d'_R\right)Z'_\mu,
\end{eqnarray}
where
\begin{eqnarray}
\Gamma_{L,u}&=&\mathcal{U}^\dagger_u\text{Diag}\left(C_{u_1,V}-C_{u_1,A},C_{u_2,V}-C_{u_2,A},C_{u_3,V}-C_{u_3,A}\right)\mathcal{U}_u,\nonumber\\
\Gamma_{R,u}&=&\mathcal{V}^\dagger_u\text{Diag}\left(C_{u_1,V}+C_{u_1,A},C_{u_2,V}+C_{u_2,A},C_{u_3,V}+C_{u_3,A}\right)\mathcal{V}_u,\nonumber\\
\Gamma_{L,d}&=&\mathcal{U}^\dagger_d\text{Diag}\left(C_{d_1,V}-C_{d_1,A},C_{d_2,V}-C_{d_2,A},C_{d_3,V}-C_{d_3,A}\right)\mathcal{U}_d,\nonumber\\
\Gamma_{R,d}&=&\mathcal{V}^\dagger_d\text{Diag}\left(C_{d_1,V}+C_{d_1,A},C_{d_2,V}+C_{d_2,A},C_{d_3,V}+C_{d_3,A}\right)\mathcal{V}_d,
\end{eqnarray}
where $C_{q_a,V}$ and $C_{q_a,A}$, with $q_a$ referring to the quark, are given in (\ref{V-A-coup}). We assume $\mathcal{U}_u=\mathcal{V}_u$ and $\mathcal{U}_d=I$ with $I$ to be the identity matrix, meaning that the presence of the Cabibbo-Kobayashi-Maskawa (CKM) quark mixing matrix $V_{\text{CKM}}$ is due to the up-type quarks.
In this sense, we have $\mathcal{U}_u=\mathcal{V}_u=V^\dagger_{\text{CKM}}$. Note that, in the following computation, the CKM matrix $V_{\text{CKM}}$ is given in the Wolfenstein parameterization as
\begin{eqnarray}
V_{\text{CKM}}=\left(%
\begin{array}{ccc}
  1-\lambda^2/2 & \lambda & A\lambda^3(\rho-i\eta) \\
  -\lambda & 1-\lambda^2/2 & A\lambda^2 \\
  A\lambda^3(1-\rho-i\eta) & -A\lambda^2 & 1 \\
\end{array}%
\right)+\mathcal{O}(\lambda^4),
\end{eqnarray}
where $\lambda\approx0.22453$, $A\approx0.836$, $\rho\approx0.12515$ and $\eta\approx0.36418$ \cite{Tanabashi2018}.

\subsection{Discrimination}

Before proceeding, let us pause here to clarify several differences between (five-dimensional) typical Kaluza-Klein model (see Ref. \cite{Bailin1987} for review) and our model.

(1) The first difference is the structure of the fundamental spacetime or the fundamental general coordinate invariance:

\begin{itemize}

\item Typical Kaluza-Klein model starts from Einstein gravity in five dimensions. This means that the initial theory has the general coordinate transformation as
\begin{eqnarray}
X^M\longrightarrow X'^M=X'^M(X^M).\label{KK-gct}
\end{eqnarray}
Then, it is assumed that because of some dynamical mechanism one of the spatial dimensions compactifies such that it has the geometry of $S^1$. In this way, the theory would reduce to the ground state $M_4\times S^1$ and the original general coordinate invariance (\ref{KK-gct}) is spontaneously broken in this ground state.

\item Whereas, our model starts from five-dimensional spacetime which is a nontrivial fiber bundle: in general, the five-dimensional spacetime is a local product of the four-dimensional manifold $M_4$ of the Lorentz signature and the Lie group manifold $\mathrm{U}(1)$ rather than a global product. In this way, the initial theory has the general coordinate transformation given by (\ref{ba-gct}).

\end{itemize}

(2) The second difference is the fundamental fermions and their action:

\begin{itemize}

\item The fermions in typical Kaluza-Klein model are the representations of the local Lorentz group $\mathrm{SO}(1,4)$. Thus, there are no the chiral fermions. Action for the fermions is given by
\begin{eqnarray}
S=\int d^5X\det(E)\left[i\bar{\psi}\gamma^A{E_A}^M\nabla_M\psi+m\bar{\psi}\psi\right],\label{KK-feract}
\end{eqnarray}
where ${E_A}^M$ are the vierbein and $\nabla_M=\partial_M+\Omega_M$ with $\Omega_M$ related to the spin connection.

\item Whereas, in our model it is possible to define the right-handed and left-handed Weyl fermions as the representations of the Lorentz group $\mathrm{SO}(1,3)$ corresponding to the horizontal metric $G_H$. This is due to the fact that it always decomposes the tangent(cotangent) space at any point of the five-dimensional spacetime into the horizontal and vertical tangent(cotangent) subspaces without reference to the local coordinate system. As a result, general action for the fermions is described by Eq. (37) in Ref. \cite{Nam2019a}. 

\end{itemize}

(3) The third difference is the charges of the SM fields under the emergent $\mathrm{U}(1)$:
\begin{itemize}
\item In typical Kaluza-Klein model, the SM fields are identified as the zero modes of the corresponding five-dimensional fields. Because of this, they are not uncharged under the emergent $\mathrm{U}(1)$ arising from the five-dimensional spacetime and thus they do not couple directly to the corresponding $\mathrm{U}(1)$ gauge field. 

\item As indicated in Ref. \cite{Nam2019a}, in our model because the SM fermions transform \emph{non-trivially} under the SM symmetry, their vertical kinetic term (which allows to determine the $\theta$-dependence of the fields) in the action should not be invariant and hence be forbidden. Thus, they themselves must have a specific $\theta$-dependence determined by their invariance under the active action of the Lie group $\textrm{U}(1)$. This leads to two consequences. First one is that the SM fermions have the emergent $\textrm{U}(1)$ charges which are the quantum numbers of characterizing the active action of the Lie group $\textrm{U}(1)$ on them. Second one is that the SM fermions have no the Kaluza-Klein counterparts, which is contrary to typical Kaluza-Klein model.
\end{itemize}

(4) The fourth difference is the stabilizing potential for the radius of the fifth dimension:

\begin{itemize}
\item In Kaluza-Klein model, there exists no the (tree level) stabilizing potential for the radius of the fifth dimension. Thus, VEV of the radion field corresponding to the fifth dimension is not stabilized and as a result the radius of the fifth dimension becomes arbitrarily large. 

\item In our model, such a stabilizing potential may be introduced at the tree level and it is given by Eq. (\ref{sta-pot}). Hence, the fifth dimension may be hidden in the low energy limit.

\end{itemize}

\section{\label{Beanomaly} $^8$Be anomaly and other constraints}
In the previous section, we have proposed a $\mathrm{U}(1)_X$ extension of the SM, which is emerged from the short-distance structure of the spacetime. There, the coupling constant $g_{_X}$ of the new gauge boson $Z'$ to the SM fermions is determined by the ratio between the inverse radius $R^{-1}$ of the fiber and the reduced four-dimensional Planck scale $M_{\text{Pl}}$ as, $g_{_X}=\frac{\sqrt{2}}{M_{\text{Pl}}R}$. This suggests that, if $R^{-1}$ is much lower than $M_{\text{Pl}}$, the coupling of $Z'$ to the SM fermions is small and $Z'$ is light but technically natural. Thus, in this section we are interested in explaining the new gauge boson $Z'$ appearing as a resonance in the Atomki experiment \cite{Krasznahorkay2016}. As reported by the Atomki Collaboration, a ratio of the branching ratios corresponding to the $^8$Be anomaly is given by 
\begin{eqnarray}
\frac{\text{BR}\left(^8\text{Be}^*\rightarrow {^8\text{Be}}+Z'\right)}{\text{BR}\left(^8\text{Be}^*\rightarrow {^8\text{Be}}+\gamma\right)}\times\text{BR}\left(Z'\rightarrow e^+e^-\right)=5.8\times10^{-6},
\end{eqnarray}
or in terms of the partial decay width as
\begin{eqnarray}
\frac{\Gamma\left(^8\text{Be}^*\rightarrow {^8\text{Be}}+Z'\right)}{\Gamma\left(^8\text{Be}^*\rightarrow {^8\text{Be}}+\gamma\right)}\times\text{BR}\left(Z'\rightarrow e^+e^-\right)=5.8\times10^{-6},
\end{eqnarray}
whose statistical significance is about $6.8\sigma$ \cite{Krasznahorkay2016}. Here, the partial decay width for $^8\text{Be}^*\rightarrow {^8\text{Be}}+\gamma$ is well-known as $\Gamma\left(^8\text{Be}^*\rightarrow {^8\text{Be}}+\gamma\right)=1.9\times10^{-6}$ MeV.

First, we need to calculate the partial decay width for $^8\text{Be}^*\rightarrow {^8\text{Be}}+Z'$ to determine what the inverse radius $R^{-1}$ of the fiber (or the coupling constant $g_{_X}$ of the gauge boson $Z'$) and the elements of the mixing matrix $\mathcal{V}_d$ mus be to explain the $^8$Be anomaly. The gauge boson $Z'$ has both vector and axial couplings to the quarks. Ref. \cite{Kozaczuk2017} showed that the contributions coming from the vector and axial couplings to the $^8\text{Be}^*\rightarrow {^8\text{Be}}+Z'$ decay are proportional to $\frac{k^3}{M^3_{Z'}}$ and $\frac{k}{M_{Z'}}$, respectively, with $k$ to be the momentum of the gauge boson $Z'$. Because $\frac{k}{M_{Z'}}\ll1$, the contribution of the axial coupling is dominant and since we can neglect the contribution of the vector coupling as well as its interference effect with the axial coupling part. We can use the result of the partial decay width for $^8\text{Be}^*\rightarrow{^8\textrm{Be}}+Z'$ mediated by an axial-vector boson in Ref. \cite{Kozaczuk2017} as
\begin{eqnarray}
  \Gamma_X &=& \frac{k}{18\pi}\left(2+\frac{E^2_k}{M^2_{Z'}}\right)\left|\frac{a_0-a_1}{2}\langle0||\sigma_n||\mathcal{S}\rangle+\frac{a_0+a_1}{2}\langle0||\sigma_p||\mathcal{S}\rangle\right|^2,
\end{eqnarray}
where $M_{Z'}=16.7$ MeV, $k=\sqrt{E^2_k-M^2_{Z'}}$, $E_k=E(^8\text{Be}^*)-E(^8\text{Be})=18.15$ MeV is the excitation energy of the isoscalar $^8\textrm{Be}^*$, the couplings of the neutron and proton are defined by
\begin{eqnarray}
  a_0 &=& \left(\Delta u^{(p)}+\Delta d^{(p)}\right)\left(C_{u,A}+C_{d,A}\right)+2C_{s,A}\Delta s^{(p)},\nonumber\\
  a_1 &=& \left(\Delta u^{(p)}-\Delta d^{(p)}\right)\left(C_{u,A}-C_{d,A}\right),
\end{eqnarray}
with the nucleon coefficients given by
\begin{equation}
\Delta u^{(p)}=0.897(27),\ \ \ \ \Delta d^{(p)}=-0.367(27),\ \ \ \ \Delta s^{(p)}=-0.026(4),
\end{equation}
and the matrix elements are given as, $\langle0||\sigma_n||\mathcal{S}\rangle=-0.132(33)$ and $\langle0||\sigma_p||\mathcal{S}\rangle=-0.047(29)$, with $|\mathcal{S}\rangle$ denoted to the isoscalar $^8\textrm{Be}^*$. In this work, the couplings $C_{u,A}$, $C_{d,A}$ and $C_{s,A}$ are given by
\begin{eqnarray}
C_{u,A}&=&\frac{\left(\Gamma_{R,u}\right)_{11}-\left(\Gamma_{L,u}\right)_{11}}{2}\simeq-\frac{g_{_X}y}{2},\nonumber\\ 
C_{d,A}&=&\frac{\left(\Gamma_{R,d}\right)_{11}-\left(\Gamma_{L,d}\right)_{11}}{2}\nonumber\\ 
&=&\frac{g_{_X}y}{2}\left[2|(\mathcal{V}_d)_{11}|^2+|(\mathcal{V}_d)_{21}|^2-1\right],\nonumber\\
C_{s,A}&=&\frac{\left(\Gamma_{R,d}\right)_{22}-\left(\Gamma_{L,d}\right)_{22}}{2}\nonumber\\
&=&-\frac{g_{_X}y}{2}\left[2-|(\mathcal{V}_d)_{22}|^2-2|(\mathcal{V}_d)_{12}|^2\right].
\end{eqnarray}
From the branching fraction of the $^8$Be anomaly and the uncertainties in the nuclear matrix elements, we can find upper and lower bounds on the inverse radius $R^{-1}$ of the fiber as
\begin{eqnarray}
\frac{1}{R}\gtrsim\frac{9.07\times|y|^{-1}\times10^{13}}{1.46|(\mathcal{V}_d)_{11}|^2-0.03|(\mathcal{V}_d)_{22}|^2+0.73|(\mathcal{V}_d)_{21}|^2-0.06|(\mathcal{V}_d)_{12}|^2-0.43}\ \ \text{GeV},\nonumber\\
\frac{1}{R}\lesssim\frac{5.71\times|y|^{-1}\times10^{14}}{3.81|(\mathcal{V}_d)_{11}|^2-0.12|(\mathcal{V}_d)_{22}|^2+1.9|(\mathcal{V}_d)_{21}|^2-0.24|(\mathcal{V}_d)_{12}|^2-2.69}\ \ \text{GeV}.\label{8Be-const}\nonumber\\
\end{eqnarray}

In the Atomki pair spectrometer, the distance between the target (where the $^8$Be nucleus is excited) and the detectors is a few cm. The requirement that the gauge boson $Z'$ must decay promptly in the detectors imposes a constraint on the electron coupling as \cite{Feng2016,Feng2017}
\begin{eqnarray}
\frac{\sqrt{C^2_{e,V}+C^2_{e,A}}}{e\sqrt{\textrm{BR}(Z'\rightarrow e^+e^-)}}\gtrsim1.3\times10^{-5}.
\end{eqnarray}
This leads to a lower bound for the inverse radius $R^{-1}$ of the fiber as
\begin{eqnarray}
\frac{1}{R}\gtrsim\frac{4.43\times10^{12}}{|y|}\ \ \text{GeV},
\end{eqnarray}
which is clearly weaker than the lower bound given in (\ref{8Be-const}).

The constraint (\ref{8Be-const}) indicates a basic requirement to explain the $^8$Be anomaly. 
In what follows, we will discuss the constraints from the various current experiments to obtain the allowed parameter space for explaining this anomaly.

Because no significant excess was observed in the decay $^8{\text{Be}^*}'\rightarrow{^8\textrm{Be}}+Z'$, where $^8{\text{Be}^*}'$ is the isovector exited state with excitation energy $17.64$ MeV, there should have a condition imposed on this decay as \cite{Feng2017,Kozaczuk2017}
\begin{eqnarray}
\frac{\Gamma\left(^8\text{Be}^*\rightarrow {^8\text{Be}}+Z'\right)}{\Gamma\left(^8\text{Be}^*\rightarrow {^8\text{Be}}+\gamma\right)}>5\times\frac{\Gamma\left(^8{\text{Be}^*}'\rightarrow {^8\text{Be}}+Z'\right)}{\Gamma\left(^8{\text{Be}^*}'\rightarrow {^8\text{Be}}+\gamma\right)},\label{excl-cond}
\end{eqnarray}
where $\Gamma\left(^8{\text{Be}^*}'\rightarrow {^8\text{Be}}+\gamma\right)=15\times10^{-6}$ MeV. The partial decay width for $^8{\text{Be}^*}'\rightarrow{^8\textrm{Be}}+Z'$ can be computed in the same way as we performed for $^8\text{Be}^*\rightarrow{^8\textrm{Be}}+Z'$, but here $E_k=E(^8{\text{Be}^*}')-E(^8\text{Be})=17.64$ MeV and the matrix elements are replaced as follows
\begin{eqnarray}
\langle0||\sigma_n||\mathcal{S}\rangle&\longrightarrow&\langle0||\sigma_n||\mathcal{V}\rangle=-0.073(29),\nonumber\\
\langle0||\sigma_p||\mathcal{S}\rangle&\longrightarrow&\langle0||\sigma_p||\mathcal{V}\rangle=0.102(28).
\end{eqnarray}
It is easily to see that the condition (\ref{excl-cond}) only places the constraint on the elements $(\mathcal{V}_d)_{11}$, $(\mathcal{V}_d)_{22}$, $(\mathcal{V}_d)_{12}$, and $(\mathcal{V}_d)_{21}$ of the mixing matrix $\mathcal{V}_d$.

The most precise measurement of the parity-violating M{\o}ller scattering from the SLAC E158 experiment \cite{Anthony2005} imposes a constraint on the coupling of the gauge boson $Z'$ to the electron as \cite{Kahn2017}
\begin{equation}
\left|C_{e,V}C_{e,A}\right|\lesssim10^{-8},\label{Moller-scat}
\end{equation}
for $M_X\approx17$ MeV. The atomic parity violation in Cesium ($^{133}_{\phantom{k} 55}\text{Cs}$) whose value is measured by the experiment \cite{Porsev09-10,Wood1997,Bennet} and is predicted by the SM \cite{Marciano84-85,Marciano90-92} as
\begin{eqnarray}
Q^{\text{exp}}_W(^{133}_{\phantom{k} 55}\text{Cs})&=&-73.16(29)_{\text{exp}}(20)_{\text{th}},\nonumber\\
Q^{\text{th}}_W(^{133}_{\phantom{k} 55}\text{Cs})&=&-73.16(3).
\end{eqnarray}
These lead to the requirement for the new physics contribution to the nuclear weak charge of Cesium as, $\left|\Delta Q_W(^{133}_{\phantom{k} 55}\text{Cs})\right|\lesssim0.52$, corresponding to the following bound on the couplings
\begin{eqnarray}
C_{e,A}\left[(2Z+N)C_{u,V}+(Z+2N)C_{d,V}\right]\lesssim6\times10^{-10},\label{APV-Ces}
\end{eqnarray}
for $M_{Z'}\approx17$ MeV where $Z=55$ and $N=78$. Clearly, the bounds (\ref{Moller-scat}) and (\ref{APV-Ces}) all are automatically satisfied because in our model we have $C_{e,A}=C_{\mu,A}=0$. 

Since the gauge boson $Z'$ has the coupling to the electron, it can be produced in the electron beam dump experiments \cite{Riordan1987,Bjorken2009}. In these experiments, the gauge boson $Z'$ could be produced in the bremsstrahlung reaction and then it would escape the beam dump and subsequently decay into the $e^+e^-$ pair. A signal of the gauge boson $Z'$ has been not observed so far in such experiments, and thus for $M_{Z'}\approx17$ MeV the SLAC E141 dump experiment puts a constraint as \cite{Rose2017,Rose2019,Kahn2017}
\begin{eqnarray}
C^2_{e,V}+C^2_{e,A}<10^{-17},\label{SLACE141-first}
\end{eqnarray}
if $Z'$ has been not produced, or 
\begin{eqnarray}
\frac{C^2_{e,V}+C^2_{e,A}}{\textrm{BR}(Z'\rightarrow e^+e^-)}\gtrsim3.7\times10^{-9},
\end{eqnarray}
if $Z'$ has been caught in the dump. Of course, the first constraint (\ref{SLACE141-first}) is not consistent to the $^8$Be anomaly and hence it is excluded. On the other hand, the coupling of the gauge boson $Z'$ to the electron is constrained by the last constraint which places a lower bound on the inverse radius $R^{-1}$ of the fiber as
\begin{eqnarray}
\frac{1}{R}\gtrsim\frac{6.81}{|y|}\times10^{13}\ \ \text{GeV}.
\end{eqnarray}
The analogous searches from Orsay \cite{Davier1989} and the SLAC E137 experiment \cite{Bjorken1988} lead to less stringent constraints on the coupling constant of the gauge boson $Z'$ or the inverse radius of the fiber. Whereas, the E774 experiment \cite{Bross1991} is only sensitive for the mass of the gauge boson $Z'$ lighter than $10$ MeV.  

Recently, collaboration of the NA64 beam dump experiment \cite{Banerjee2018} reported the first results in attempting to search for the hypothetical $17$ MeV gauge boson. This showed that, for the explanation of the gauge boson $Z'$ as the $^8$Be anomaly, its coupling to the electron must satisfy the following constraint
\begin{eqnarray}
\frac{C^2_{e,V}+C^2_{e,A}}{\textrm{BR}(Z'\rightarrow e^+e^-)}\gtrsim1.6\times10^{-8},
\end{eqnarray}
which leads to a lower bound on the inverse radius of the fiber as
\begin{eqnarray}
\frac{1}{R}\gtrsim\frac{1.42}{|y|}\times10^{14}\ \ \text{GeV}.\label{NA64-bound}
\end{eqnarray} 
It is clearly that this lower bound is stronger than one placed by the SLAC E141 dump experiment. 

Also, the gauge boson $Z'$ could be emitted in the decays of the $\eta$ and $\eta'$ neutral mesons which are produced by the high energy proton beam in a neutrino target \cite{Gninenko2012}. Using the constraints from search for the signature of the heavy neutrino decay $\nu_h\rightarrow\nu e^+e^-$, the CHARM experiment at CERN puts a bound as, $\left(C^2_{e,V}+C^2_{e,A}\right)/\text{Br}(Z'\rightarrow e^+e^-)\gtrsim3.7\times10^{-11}$ for $M_{Z'}\approx17$ MeV \cite{Feng2017}, which is clearly much weaker than the bound of the NA64 experiment.

Furthermore, the coupling of the gauge boson $Z'$ to the electron is constrained by the electron-positron colliding experiment, e.g. KLOE2 \cite{Babusci2013}. The search for the process $e^+e^-\rightarrow\gamma(Z'\rightarrow e^+e^-)$ has set a constraint as, $\left(C^2_{e,V}+C^2_{e,A}\right)\text{Br}(Z'\rightarrow e^+e^-)\lesssim3.7\times10^{-7}$ for $M_{Z'}\approx17$ MeV \cite{Rose2017,Rose2019,Kahn2017}, which leads to an upper bound on the inverse radius of the fiber as
\begin{eqnarray}
\frac{1}{R}\lesssim\frac{6.81}{|y|}\times10^{14}\ \ \text{GeV}.
\end{eqnarray} 
An analogous search, e.g. the BABAR experiment \cite{Lees2014}, is only sensitive to the mass of the gauge boson $Z'$ heavier than $20$ MeV.

In addition, the precise measurement of the mass of the SM boson $Z$ which is $M_Z=91.1876\pm0.0021$ GeV \cite{Tanabashi2018} leads to 
\begin{eqnarray}
|\Delta M_Z|=\left(M_Z-\frac{\sqrt{g^2+g'^2}}{2}v\right)\simeq\frac{g^2_{_X}y^2v}{\sqrt{g^2+g'^2}}\lesssim0.0021\ \ \text{GeV},
\end{eqnarray}
This corresponds to the following upper bound
\begin{equation}
\frac{1}{R}\lesssim\frac{4.33}{|y|}\times10^{15}\ \ \text{GeV}.
\end{equation}

Because the gauge boson $Z'$ is coupled to the charged leptons, it should contribute to the anomalous magnetic moments of the electron and muon. In our model, the gauge boson $Z'$ has only the vector couplings to the electron and muon. Thus, the one-loop contributions for the anomalous magnetic moments of the electron and muon, mediated by $Z'$, are given by
\begin{eqnarray}
\delta a_e&=&\frac{C^2_{e,V}}{4\pi^2}\int_{0}^{1}dz\frac{z^2(1-z)}{\frac{M^2_{Z'}}{m^2_e}(1-z)+z^2}\simeq\left(g_{_X}y\right)^2\times1.84\times10^{-5},\nonumber\\
\delta a_\mu&=&\frac{C^2_{\mu,V}}{4\pi^2}\int_{0}^{1}dz\frac{z^2(1-z)}{\frac{M^2_{Z'}}{m^2_\mu}(1-z)+z^2}\simeq\left(g_{_X}y\right)^2\times2.04\times10^{-2}.
\end{eqnarray}
Now we use the discrepancy between the measured values of the anomalous magnetic moments of the electron and muon and the SM predictions to impose the constraints on the coupling of the gauge boson $Z'$ or the inverse radius $R^{-1}$ of the fiber. For the anomalous magnetic moments of the electron, the new contributions must satisfy $-26\times10^{-13}\lesssim\delta a_e\lesssim8\times10^{-13}$ \cite{Giudice2012} which leads to the following bound
\begin{eqnarray}
\frac{1}{R}\lesssim\frac{3.6}{|y|}\times10^{14}\ \ \text{GeV}.\label{EAMM}
\end{eqnarray}
For the anomalous magnetic moments of the muon, the positive discrepancy between the measurement and the SM prediction is about $\delta a_\mu=a^{\text{exp}}_\mu-a^{\text{th}}_\mu=306\pm72\times10^{-11}$
which is at $4.3\sigma$ \cite{Jegerlehner2017,Bennett2006}. Since the new contribution coming from the gauge boson $Z'$ is required to satisfy $\delta a_\mu\lesssim3.78\times10^{-9}$ in $90\%$ C.L. from which we obtain an upper bound as
\begin{eqnarray}
\frac{1}{R}\lesssim\frac{7.4}{|y|}\times10^{14}\ \ \text{GeV}.
\end{eqnarray}
Note that, in the scenarios in which the new neutral gauge boson has both axial and vector couplings to the muon, the corresponding one-loop contribution is given about $0.009C^2_{\mu,V}-C^2_{\mu,A}$ for $M_{Z'}\approx17$ MeV which is negative and thus disagrees \cite{Kozaczuk2017,Rose2017}. On the other hand, the constraint is imposed on the axial coupling whose contribution is required to be less than the $2\sigma$ uncertainty of the discrepancy between the measurement and the SM prediction.

We now analyze the constraints from the couplings of the gauge boson $Z'$ to the quarks. The axial $Z'$ couplings would contribute to the amplitude of the rare $\eta$ decay $\eta\rightarrow\mu^+\mu^-$. This imposes the constraint on the product $C_{\mu,A}\left(C_u+C_d-cC_s\right)$ where $C_f=\sqrt{C^2_{f,V}+C^2_{f,A}}$ and $c$ is the parameter relating to the $\eta$-$\eta'$ mixing. In our model, this constraint is avoided because $C_{\mu,A}=0$. Also, the rare neutral pion decay $\pi^0\rightarrow\gamma(Z'\rightarrow e^+e^-)$ from the NA48/2 experiment \cite{Raggi2016} imposes the constraint on the coupling of the gauge boson $Z'$ to the first generation of the quarks. The contribution coming from the axial couplings is suppressed by the masses of the light quarks. Since it is mainly imposed the bound on the vector couplings as \cite{Rose2019}
\begin{equation}
\left|2C_{u,V}+C_{d,V}\right|\lesssim\frac{3.6\times10^{-4}}{\sqrt{\textrm{BR}(Z'\rightarrow e^+e^-)}},
\end{equation}
for $M_{Z'}\approx17$ MeV. This leads to the following upper bound on the inverse radius of the fiber as
\begin{eqnarray}
\frac{1}{R}\lesssim\frac{5.44\times|y|^{-1}\times10^{14}}{0.12+0.88\left[|(\mathcal{V}_d)_{11}|^2+|(\mathcal{V}_d)_{21}|^2/2\right]}\ \ \text{GeV}.\label{NA48/2}
\end{eqnarray}
Furthermore, the measurements of the neutron-nucleus scattering set a bound on the combination of the $Z'$ couplings of the up and down quarks as $\left(2C_d+C_u\right)^2\lesssim13.6\times\pi\times10^{-11}\times\left(M_{Z'}/\text{MeV}\right)^4$ \cite{Barbieri1975}. Clearly, this bound is weaker than the bound obtaining from the rare neutral pion decay. There is additionally the constraint on the $Z'$ couplings to the second generation of the quarks. The decay of $\phi\rightarrow\eta(Z'\rightarrow e^+e^-)$ sets a bound on the $Z'$ coupling of the strange quark as, $|C_s|\sqrt{\text{Br}(Z'\rightarrow e^+e^-)}\lesssim1.0\times10^{-2}$ for $M_{Z'}\approx17$ MeV \cite{Babusci2013}. In the case $C_s\approx C_d$, this constraint is is much weaker than the constraint obtaining from the rare neutral pion decay.

The gauge boson $Z'$ has the flavor-nondiagonal couplings to quarks and thus it would lead to
the quark-flavor violating processes at the tree-level. As a result, the $Z'$ couplings to the quarks (or the inverse radius of the fiber) and the mixing matrix $\mathcal{V}_d$ should be constrained by the relevant experimental bounds. With $M_{Z'}\approx17$ MeV, the gauge boson $Z'$ would contribute to the decay of the mesons, such as $K^0\rightarrow\pi^0e^+e^-$ or $B^0\rightarrow K^{*0}e^+e^-$. However, in search for such decays in LHCb, the $e^+e^-$ invariant mass is above $20$ MeV \cite{Aaij2015,Aaij2017}. And, since the constraints on the $Z'$ couplings to the quarks and the mixing matrix $\mathcal{V}_d$ from these decays require the future upgrades of the LHCb experiment.

Now we study the contribution of the gauge boson $Z'$ to the $\Delta F=2$ transition or the mixing of the neutral meson systems. The effective Lagrangian, that describes the mixing of the neutral meson systems mediated by the gauge boson $Z'$, is given by
\begin{eqnarray}
\mathcal{L}^{Z'}_{\text{eff}}[q_i,q_j]&=&-\left[\frac{C^{LL}_{ij}(q^2)}{2}\left(\bar{q}_{iL}\gamma_\mu q_{jL}\right)\left(\bar{q}_{iL}\gamma^\mu q_{jL}\right)+\frac{C^{RR}_{ij}(q^2)}{2}\left(\bar{q}_{iR}\gamma_\mu q_{jR}\right)\left(\bar{q}_{iR}\gamma^\mu q_{jR}\right)\right.\nonumber\\
&&\left.+ C^{LR}_{ij}(q^2)\left(\bar{q}_{iL}\gamma_\mu q_{jL}\right)\left(\bar{q}_{iR}\gamma^\mu q_{jR}\right)+\textrm{H.c.}\right],
\end{eqnarray}
where 
\begin{eqnarray}
C^{LL}_{ij}(q^2)&=&\frac{\left(\Gamma_{L,q}\right)^2_{ij}}{q^2-M^2_{Z'}},\nonumber\\
C^{RR}_{ij}(q^2)&=&\frac{\left(\Gamma_{R,q}\right)^2_{ij}}{q^2-M^2_{Z'}},\nonumber\\
C^{LR}_{ij}(q^2)&=&\frac{\left(\Gamma_{L,q}\right)_{ij}\left(\Gamma_{R,q}\right)_{ij}}{q^2-M^2_{Z'}},
\end{eqnarray}
and $q^2$ refers to the momentum transfer. It should be noted here that, for the down-type quarks in our model, $C^{LL}_{ij}(q^2)=0$ with $i\neq j$. With this effective Langrangian, one can find the mass difference for the mixing of the $S-\bar{S}$ meson system, as
\begin{eqnarray}
\Delta M_{S}&=&-2\text{Re}\langle\bar{S}|\mathcal{L}^{Z'}_{\text{eff}}[q_i,q_j]|S\rangle,\nonumber\\
&=&\frac{2M_Sf^2_S}{3}\text{Re}\left[C^{LL}_{ij}(q^2)+C^{RR}_{ij}(q^2)-2C^{LR}_{ij}(q^2)\right].
\end{eqnarray}
Using the results in Ref. \cite{Blum2009} for the  $K^0-\bar{K}^0$ and $D^0-\bar{D}^0$ mixings, we can place the corresponding bounds as
\begin{eqnarray}
\Big|C^{RR}_{ds}(M^2_K)\Big|\lesssim\frac{8.8\times10^{-7}}{\text{TeV}^2},\nonumber\\
\Big|C^{LL}_{uc}(M^2_D)+C^{RR}_{uc}(M^2_D)-2C^{LR}_{uc}(M^2_D)\Big|\lesssim\frac{5.9\times10^{-7}}{\text{TeV}^2}.
\end{eqnarray}
This leads to
\begin{eqnarray}
\frac{1}{R}\lesssim\frac{2.41\times|y|^{-1}\times10^{12}}{\Big|(\mathcal{V}_d)_{12}-2(\mathcal{V}^*_d)_{21}+2s^2_W\left[(\mathcal{V}_d)_{12}+(\mathcal{V}^*_d)_{21}\right]\Big|}\ \ \text{GeV},\label{KKmixing}\\
\frac{1}{R}\lesssim\frac{8.03}{|y|}\times10^{15}\ \ \text{GeV}.
\end{eqnarray} 
Note that, we have taken approximately the diagonal elements of the mixing matrix $\mathcal{V}_d$ to be equal to one and we have neglected the quadratic term in the off-diagonal elements, because the off-diagonal elements of $\mathcal{V}_d$ are small due to the constraints. Experimental and SM values for the mass difference in the  $B_d-\bar{B}_d$ and $B_s-\bar{B}_s$ meson systems are given by \cite{King2019}
\begin{eqnarray}
\Delta M^{\text{exp}}_{B_d}&=&(0.5064\pm0.0019)\ \ \text{ps}^{-1},\ \ \ \ \ \ \Delta M^{\text{SM}}_{B_d}=(0.547^{+0.035}_{-0.046})\ \ \text{ps}^{-1},\nonumber\\
\Delta M^{\text{exp}}_{B_s}&=&(17.757\pm0.021)\ \ \text{ps}^{-1},\ \ \ \ \ \ \Delta M^{\text{SM}}_{B_s}=(18.5^{+1.2}_{-1.5})\ \ \text{ps}^{-1}.
\end{eqnarray}
From this, we can obtain the $2\sigma$ lower bounds on the $B_d-\bar{B}_d$ and $B_s-\bar{B}_s$ mixings as, $\Delta M^{\text{SM}}_{B_d}>0.455\ \ \text{ps}^{-1}$ and $\Delta M^{\text{SM}}_{B_s}>15.5\ \ \text{ps}^{-1}$, respectively. For the $B_d-\bar{B}_d$ mixing, the contribution of the gauge boson $Z'$ satisfies the $2\sigma$ bound as
\begin{eqnarray}
\frac{1}{R}\lesssim\frac{1.43\times|y|^{-1}\times10^{13}}{\left(\text{Re}\left\{(\mathcal{V}_d)_{13}-5(\mathcal{V}^*_d)_{31}+2s^2_W\left[(\mathcal{V}_d)_{13}+(\mathcal{V}^*_d)_{31}\right]\right\}^2\right)^{1/2}}\ \ \text{GeV}.\label{Bdmixing}
\end{eqnarray}
Whereas, for the $B_s-\bar{B}_s$ mixing, the $2\sigma$ bound is given by
\begin{eqnarray}
\frac{1}{R}\lesssim\frac{7.8\times|y|^{-1}\times10^{13}}{\left(\text{Re}\left\{2(\mathcal{V}_d)_{23}+5(\mathcal{V}^*_d)_{32}-2s^2_W\left[(\mathcal{V}_d)_{23}+(\mathcal{V}^*_d)_{32}\right]\right\}^2\right)^{1/2}}\ \ \text{GeV}.\label{Bsmixing}
\end{eqnarray}
We can see that the bound coming from the $D^0-\bar{D}^0$ mixing almost satisfies the requirement to explain the $^8$Be anomaly. Whereas, for the remaining bounds, the bound coming from the $K^0-\bar{K}^0$ mixing is the most stringent one if the amplitudes of the off-diagonal elements of $\mathcal{V}_d$ are approximately equal together. 

In summary, combining all the required bounds obtained above, we have the allowed parameter space to explain the $^8$Be anomaly. The allowed parameter space is essentially determined by the basic requirement (\ref{8Be-const}), the constraint (\ref{excl-cond}) corresponding to no significant excess observed in the $^8{\text{Be}^*}'$ decay, the NA64 bound (\ref{NA64-bound}), the $(g-2)_e$ bound (\ref{EAMM}), the NA48/2 bound (\ref{NA48/2}), the bounds (\ref{KKmixing}), (\ref{Bdmixing}), and (\ref{Bsmixing}) from the mixing of the neutral meson systems. In the following analysis, we will take $(\mathcal{V}_d)_{11},(\mathcal{V}_d)_{22}, (\mathcal{V}_d)_{33}\approx1$, $(\mathcal{V}_d)_{12}=-(\mathcal{V}_d)_{21}$, $(\mathcal{V}_d)_{13}=-(\mathcal{V}_d)_{31}$, and $(\mathcal{V}_d)_{23}=-(\mathcal{V}_d)_{32}$. 

\begin{figure}[t]
% Requires \usepackage{graphicx}\
 \centering
\begin{tabular}{cc}
\includegraphics[width=0.47 \textwidth]{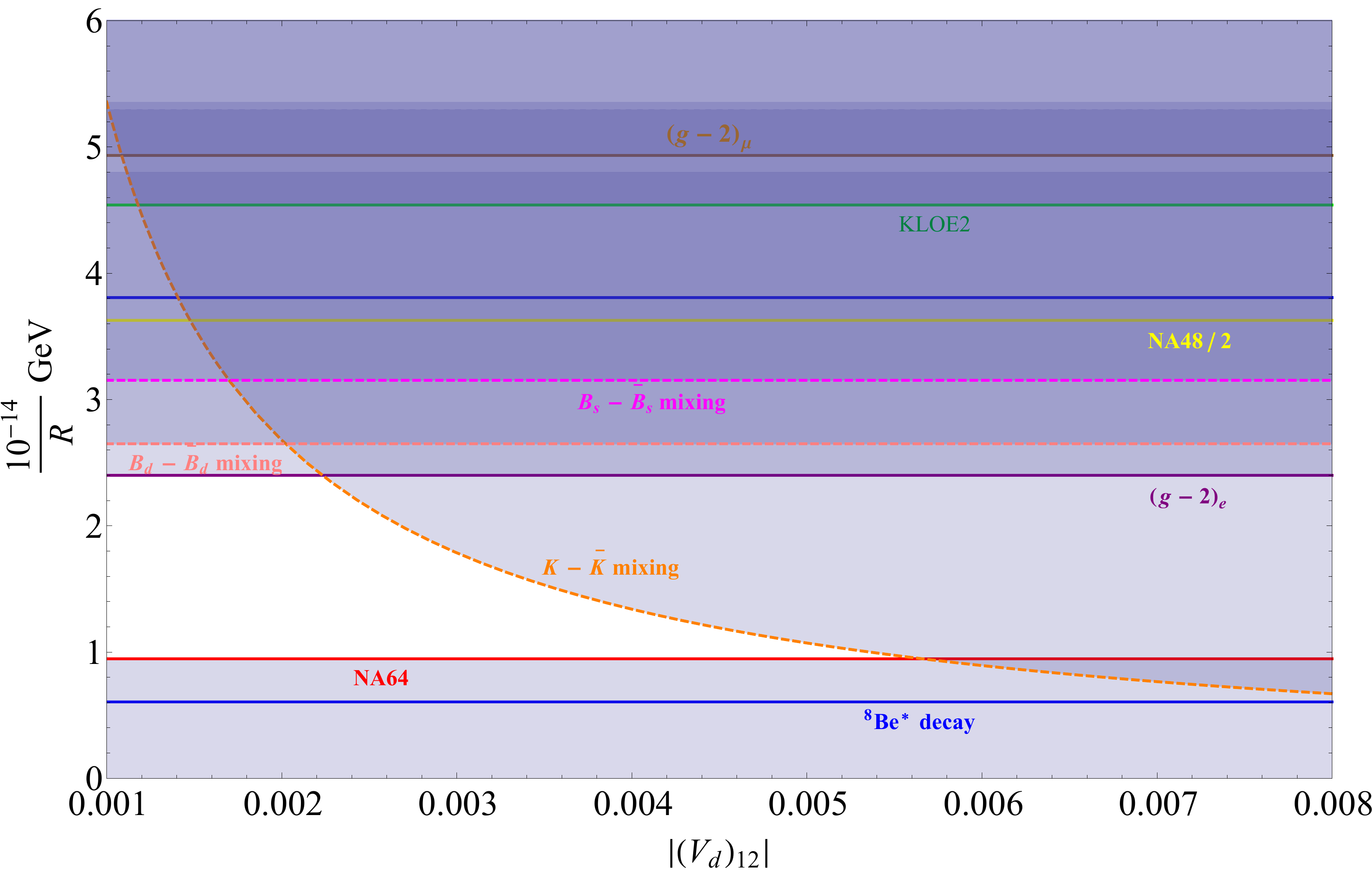}
\hspace*{0.01\textwidth}
\includegraphics[width=0.47 \textwidth]{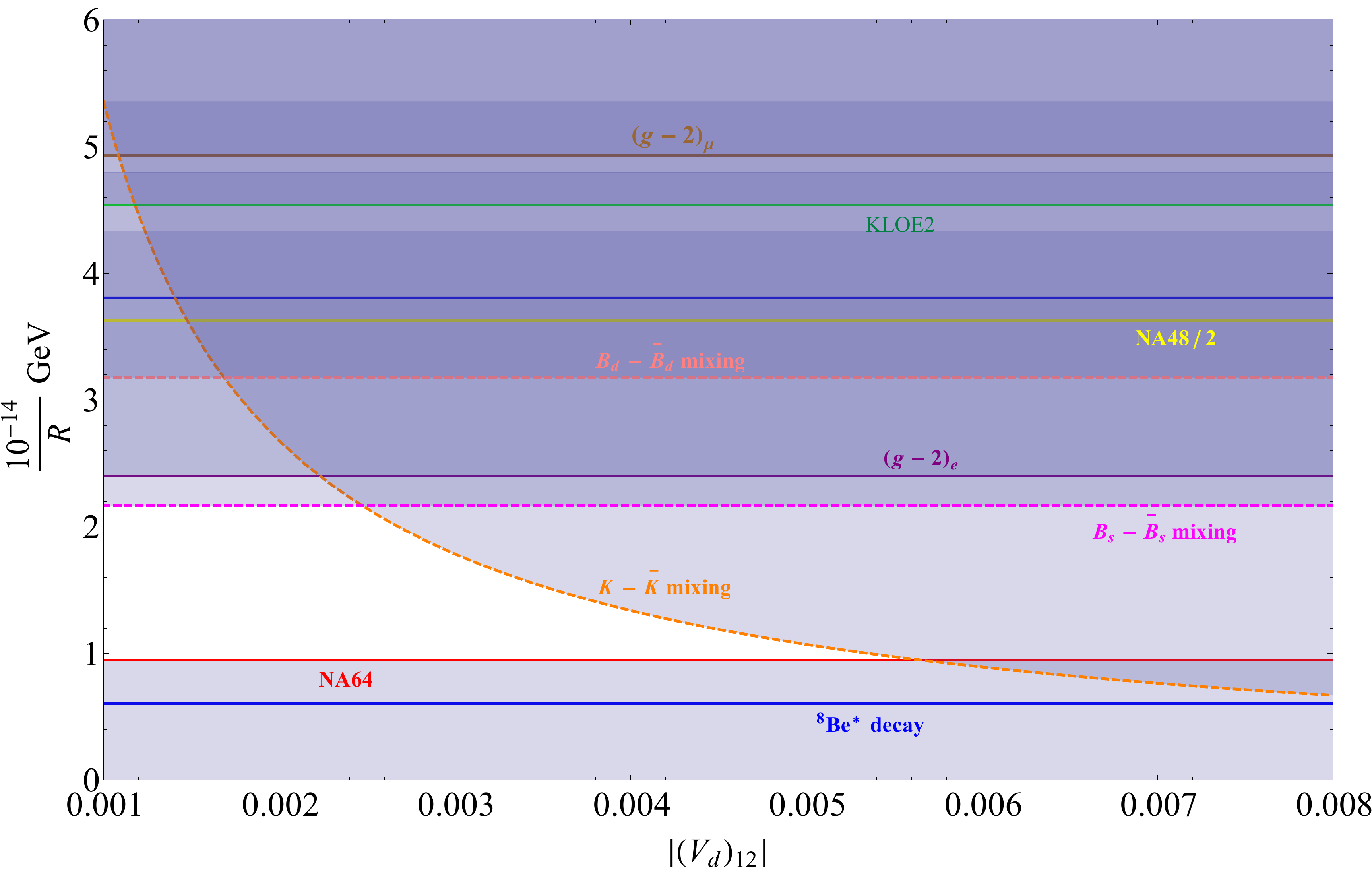}
\end{tabular}
  \caption{The possibly allowed parameter space (white region) for the inverse radius $1/R$ of the fiber and the element $|(\mathcal{V}_d)_{12}|$, at $y=3/2$. Left panel: $(\mathcal{V}_d)_{13}=0.006$ and $(\mathcal{V}_d)_{23}=0.055$. Right panel: $(\mathcal{V}_d)_{13}=0.005$ and $(\mathcal{V}_d)_{23}=0.08$.}\label{para-space}
\end{figure}

In fact, it is difficult to find a bound corresponding to the constraint (\ref{excl-cond}) without introducing the uncertainties in the nuclear matrix elements and nucleon coefficients. Therefore, first we use the bounds mentioned above except (\ref{excl-cond}) to find the possibly allowed parameter space. Then, with given uncertainties in the nuclear matrix elements and nucleon coefficients, one can obtain the bound corresponding to (\ref{excl-cond}) and thus determine the points of that region which are consistent or not to the constraint (\ref{excl-cond}). In Fig. \ref{para-space}, we plot the possibly allowed parameter space, given by the white region, in the plane of the inverse radius $R^{-1}$ of the fiber and the element $|(\mathcal{V}_d)_{12}|$ of the mixing matrix $\mathcal{V}_d$.
The possibly allowed range of the gauge coupling constant $g_{_X}$ is given by $5.5\times10^{-5}\lesssim g_{_X}\lesssim1.4\times10^{-4}$ and $5.5\times10^{-5}\lesssim g_{_X}\lesssim1.2\times10^{-4}$ corresponding to the left and right panels of Fig. \ref{para-space}, respectively. With $\langle0||\sigma_n||\mathcal{S}\rangle=-0.132-0.01$, $\langle0||\sigma_p||\mathcal{S}\rangle=-0.047$, $\langle0||\sigma_n||\mathcal{V}\rangle=-0.073 + 0.02$, and $\langle0||\sigma_p||\mathcal{V}\rangle=0.102 - 0.02$, we can get $g_{_X}\simeq5.6\times10^{-5}$ which belongs the possibly allowed range and the constraint (\ref{excl-cond}) becomes
\begin{eqnarray}
0.0244+0.2024|(\mathcal{V}_d)_{12}|^2\gtrsim0,
\end{eqnarray}
which is always satisfied for any value of $(\mathcal{V}_d)_{12}$. Thus the points with $g_{_X}\simeq5.6\times10^{-5}$ in the white region of Fig. \ref{para-space} can allow to explain $^8\text{Be}$ anomaly. Whereas, with $\langle0||\sigma_n||\mathcal{S}\rangle=-0.132$, $\langle0||\sigma_p||\mathcal{S}\rangle=-0.047$, $\langle0||\sigma_n||\mathcal{V}\rangle=-0.073+0.01$, and $\langle0||\sigma_p||\mathcal{V}\rangle=0.102-0.01$, we can get $g_{_X}\simeq9.4\times10^{-5}$ which belongs the possibly allowed range and the constraint (\ref{excl-cond}) becomes
\begin{eqnarray}
-0.1252+0.1346|(\mathcal{V}_d)_{12}|^2\gtrsim0 \ \ \ \ \longrightarrow \ \ \ \ |(\mathcal{V}_d)_{12}|\gtrsim0.96.
\end{eqnarray}
This means that the points with $g_{_X}\simeq9.4\times10^{-5}$ in the white region of Fig. \ref{para-space} can not explain $^8\text{Be}$ anomaly. Note that, the nucleon coefficients $\Delta u^{(p)}$, $\Delta d^{(p)}$, and $\Delta s^{(p)}$ have been taken at the corresponding center values.

\begin{figure}[t]
% Requires \usepackage{graphicx}\
 \centering
\begin{tabular}{cc}
\includegraphics[width=0.47 \textwidth]{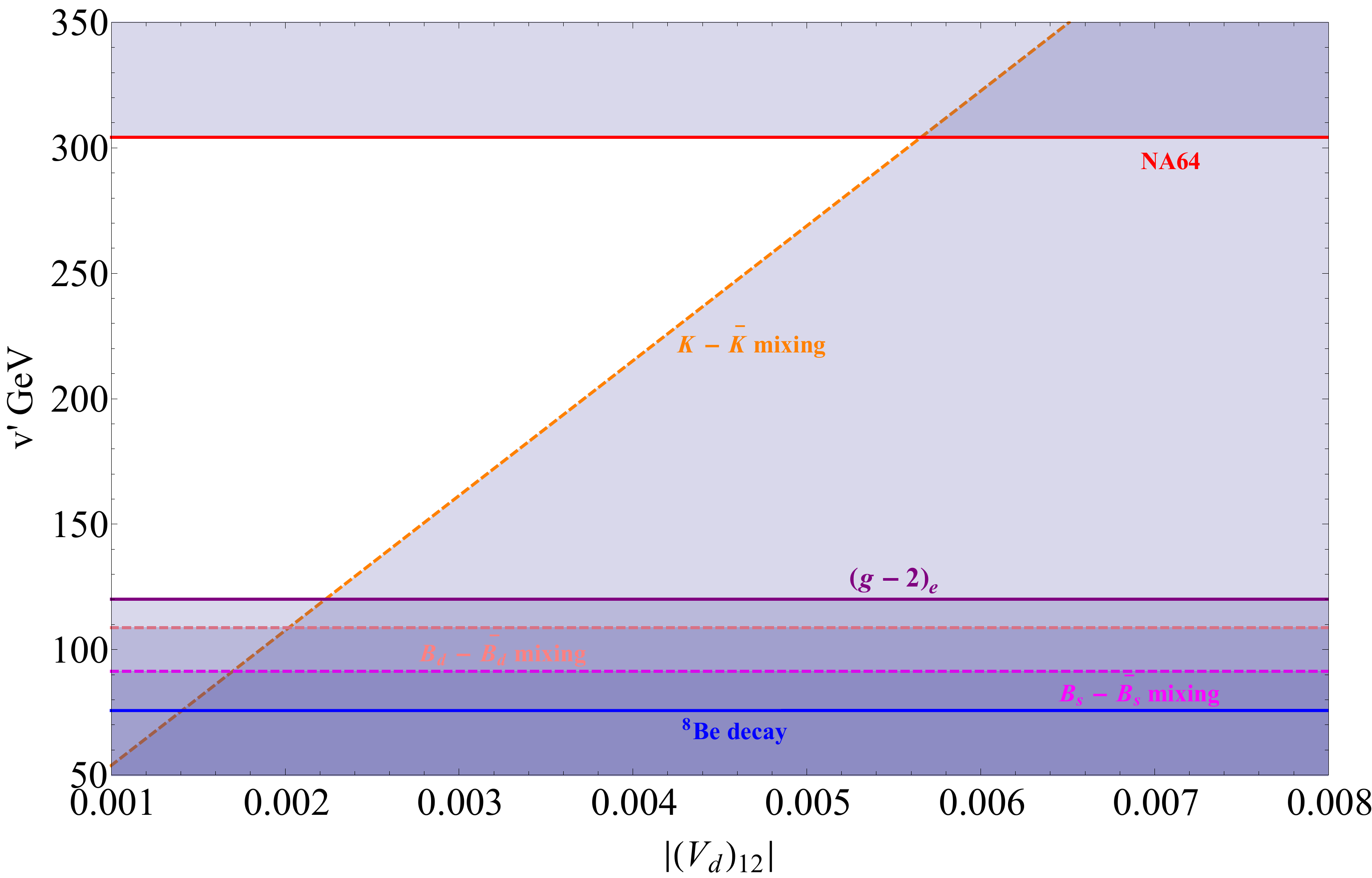}
\hspace*{0.01\textwidth}
\includegraphics[width=0.47 \textwidth]{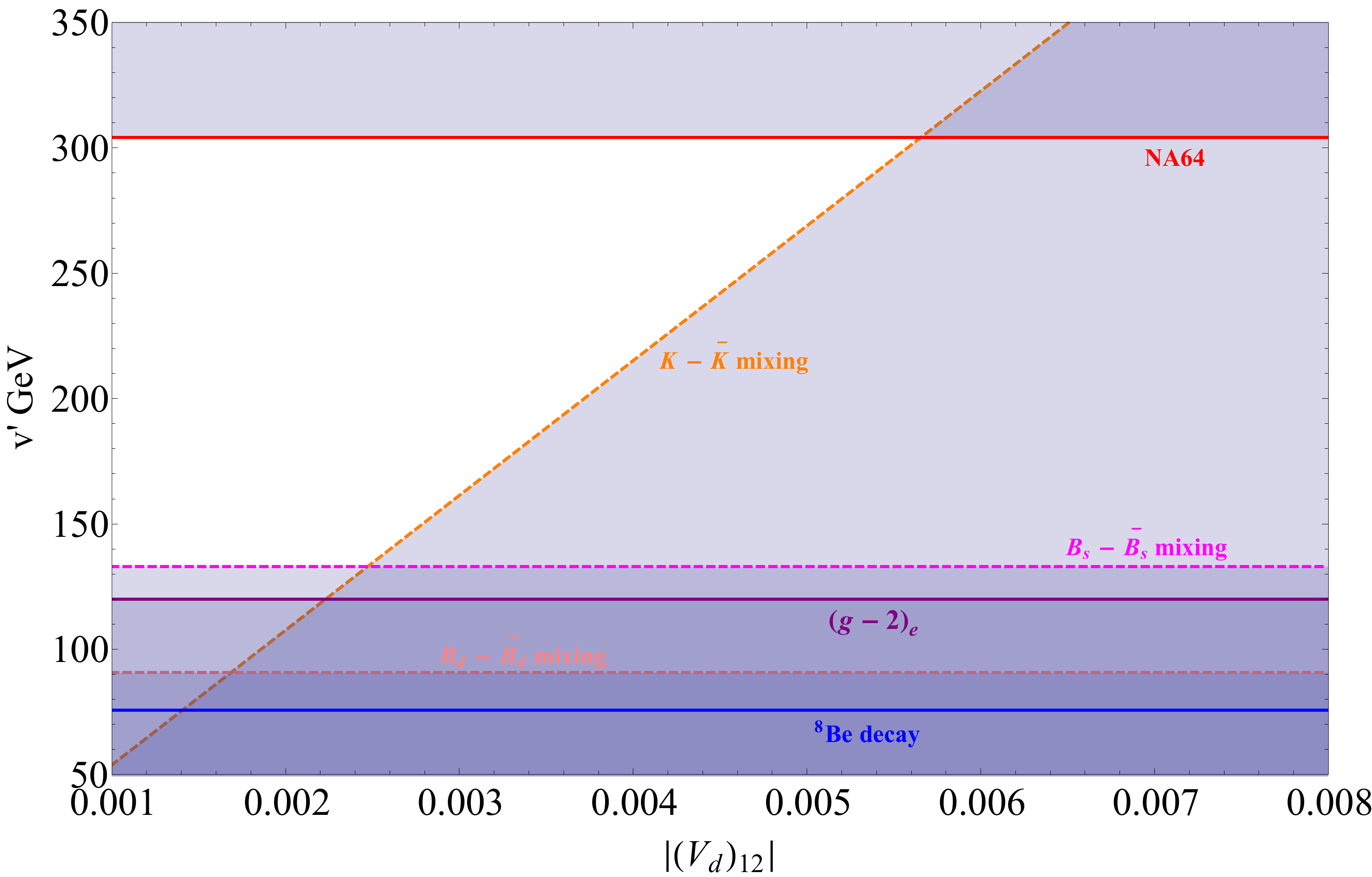}
\end{tabular}
  \caption{The possibly allowed parameter space (white region) for VEV $v'$ of the exotic Higgs field $\varphi$ and the element $|(\mathcal{V}_d)_{12}|$, at $y=3/2$. Left panel: $(\mathcal{V}_d)_{13}=0.006$ and $(\mathcal{V}_d)_{23}=0.055$. Right panel: $(\mathcal{V}_d)_{13}=0.005$ and $(\mathcal{V}_d)_{23}=0.08$.}\label{VEVv2-par-space}
\end{figure}

From the expression for the squared masses of the SM neutral boson $Z$ and the new neutral one $Z'$, given in (\ref{neut-bos-mass}), we find the following relation for the VEV $v'$ of the exotic Higgs field $\varphi$ as
\begin{equation}
v'=\frac{2M_ZM_{Z'}}{g_{_X}\sqrt{g^2+g'^2}v}\simeq2.88\times10^{16}\times\frac{\text{GeV}^2}{R^{-1}}.\label{VEVv2-inrad}
\end{equation}
In Fig. \ref{VEVv2-par-space}, we plot the possibly allowed parameter space, which is specified by the white region, in the plane of the VEV $v'$ and the element $|(\mathcal{V}_d)_{12}|$ of the mixing matrix $\mathcal{V}_d$.
In similar way as above, we can determine the points, in the white region, which are consistent or not to the constraint (\ref{excl-cond}) with given uncertainties in the nuclear matrix elements.

\begin{figure}[t]
% Requires \usepackage{graphicx}\
 \centering
\begin{tabular}{cc}
\includegraphics[width=0.47 \textwidth]{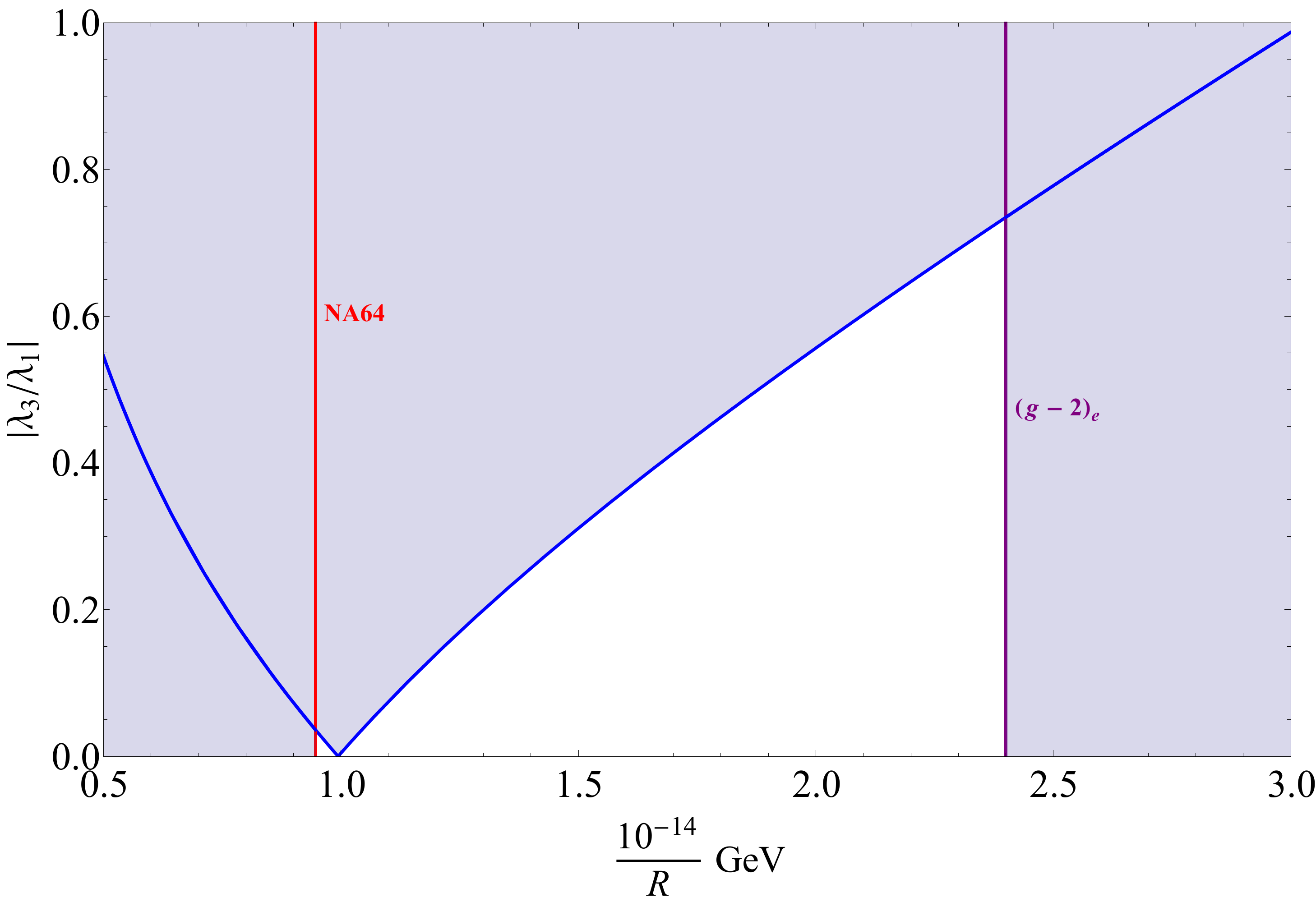}
\hspace*{0.01\textwidth}
\includegraphics[width=0.47 \textwidth]{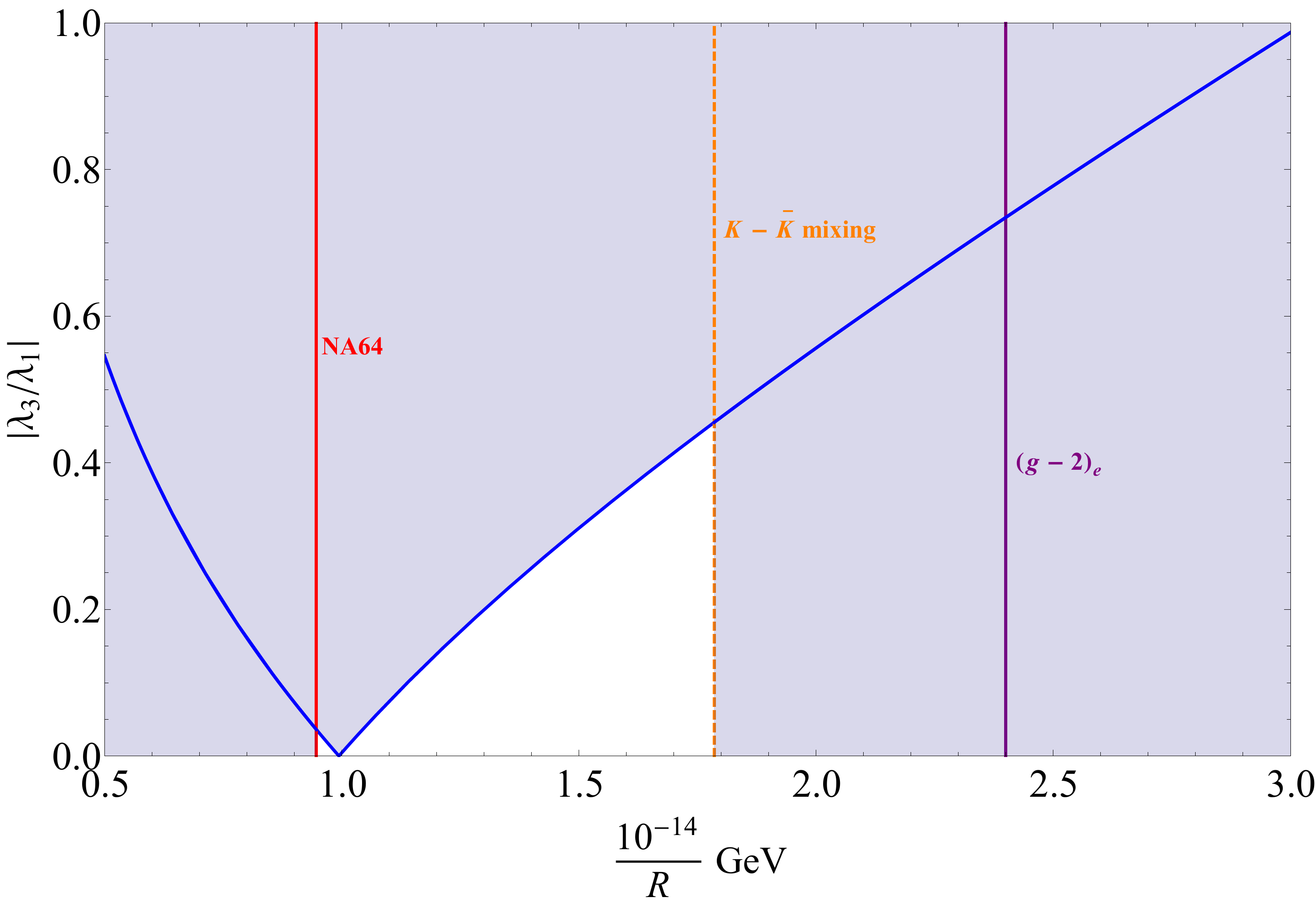}
\end{tabular}
  \caption{The possibly allowed parameter space (white region) for $|\lambda_3/\lambda_1|$ and the inverse radius $1/R$ of the fiber, at $y=3/2$, $(\mathcal{V}_d)_{13}=0.006$ and $(\mathcal{V}_d)_{23}=0.055$. Left panel: $(\mathcal{V}_d)_{12}=0.002$. Right panel: $(\mathcal{V}_d)_{12}=0.003$. The blue lines correspond to the constraint (\ref{l3l1-constr}).}\label{l3l1-par-space}
\end{figure}

Furthermore, using the bound (\ref{lam3-constr}) and  the relation (\ref{VEVv2-inrad}), one can obtain the constraint on $|\lambda_3/\lambda_1|$ as
\begin{eqnarray}
\left|\frac{\lambda_3}{\lambda_1}\right|&\lesssim&\left|0.63 g_{_X}-\frac{3\times10^{-9}}{g_{_X}}\right|\times10^4,\nonumber\\
&\lesssim&\left|\frac{9R^{-1}}{M_{Pl}}-\frac{3M_{Pl}}{2R^{-1}}\times10^{-8}\right|\times10^3,\label{l3l1-constr}
\end{eqnarray}
for $\lambda_1\approx\lambda_2$. The possibly allowed parameter space in the plane of $|\lambda_3/\lambda_1|$ and $R^{-1}$ is shown in Fig. \ref{l3l1-par-space}.
Then, the points in the white region, which are consistent or not to the constraint (\ref{excl-cond}), can be determined with given uncertainties in the nuclear matrix elements.

\section{\label{conclu} Conclusion}

The Atomki pair spectrometer experiment has recently reported an anomaly in the $^8$Be nuclear transition, which hints at a new weakly-coupled, light neutral gauge boson. Motivated by this, many theoretical models have been constructed to explain this anomaly, at which the new gauge boson comes from the extension of the SM gauge symmetry group through adding an extra $\mathrm{U}(1)$ group(s). In this work, based on our recent work \cite{Nam2019a}, we proposed a realistic model where the new gauge boson arises from a short-distance structure of the spacetime. We departs fundamentally from a generally covariant theory in the five-dimensional fiber bundle spacetime with the Standard Model (SM) gauge symmetry. We obtained in the four-dimensional effective spacetime an extension of the SM with the gauge symmetry $\mathrm{SU}(3)_C\otimes\mathrm{SU}(2)_L\otimes\mathrm{U}(1)_Y\otimes\mathrm{U}(1)_X$. We identify the new gauge boson as the physical state which mixes between the SM neutral boson and the gauge boson corresponding to $\mathrm{U}(1)_X$. Interestingly, the coupling constant of the new gauge boson is naturally defined by the ratio between the inverse radius of the fiber and the reduced four-dimensional Planck scale. We determined the required inverse radius of the fiber, which is in order of $\mathcal{O}\left(10^{14}\right)$ GeV, to explain the $^8$Be anomaly and satisfy the constraints from various experiments

\section*{Acknowledgments}
We would like to express sincere gratitude to the referee for his constructive comments and suggestions which have helped us to improve the quality of the paper.

\end{document}